\documentclass[aps,pra,reprint,a4paper,showpacs,superscriptaddress,floatfix,longbibliography]{revtex4-1}
\usepackage[pdftex]{graphicx}% Include figure files
\usepackage{dcolumn}% Align table columns on decimal point
\usepackage{bm}% bold math
\usepackage[usenames, dvipsnames]{color}
\usepackage{comment}

\usepackage[T1]{fontenc}

\usepackage{colonequals}
\usepackage{amsfonts}
\usepackage{amssymb}
\usepackage{amsmath}
\usepackage{amsthm}
\usepackage{bbm}
\usepackage{multirow}
\usepackage[colorinlistoftodos]{todonotes}
\usepackage[scr=boondoxo, scrscaled=1.05]{mathalfa}

\usepackage{subfigure}

\usepackage{letltxmacro}
\makeatletter
\let\oldr@@t\r@@t
\def\r@@t#1#2{%
\setbox0=\hbox{$\oldr@@t#1{#2\,}$}\dimen0=\ht0
\advance\dimen0-0.2\ht0
\setbox2=\hbox{\vrule height\ht0 depth -\dimen0}%
{\box0\lower0.4pt\box2}}
\LetLtxMacro{\oldsqrt}{\sqrt}
\renewcommand*{\sqrt}[2][\ ]{\oldsqrt[#1]{#2}}
\makeatother

\usepackage{array}
\newcolumntype{L}[1]{>{\raggedright\let\newline\\\arraybackslash\hspace{0pt}}m{#1}}
\newcolumntype{C}[1]{>{\centering\let\newline\\\arraybackslash\hspace{0pt}}m{#1}}
\newcolumntype{R}[1]{>{\raggedleft\let\newline\\\arraybackslash\hspace{0pt}}m{#1}}
\usepackage{tabularx}

\def\KeyWord#1{$\backslash$\IfColor{$\!\!$\textRed{#1}\textBlack}{#1}$\!\!$}

\def\red#1{#1}
\def\redd#1{#1}
\def\para#1{\textit{\red{Aim of para: #1}}\par}
\def\para#1{}

\usepackage{wasysym}

\newcommand{\be}{\begin{equation} }
\newcommand{\ee}{\end{equation} }
\newcommand{\ba}{\begin{eqnarray} }
\newcommand{\ea}{\end{eqnarray} }
\newcommand{\bit}{\begin{itemize}}
\newcommand{\eit}{\end{itemize}}
\newcommand{\ben}{\begin{enumerate}}
\newcommand{\een}{\end{enumerate}}

\renewcommand{\d}{\mathrm{d}}
\newcommand{\e}{\mathrm{e}}

\def\bra#1{\langle#1|}
\def\ket#1{|#1\rangle}

\def\tr#1{\mathrm{tr}\left(#1\right)}
\def\ptr#1#2{\mathrm{tr}_{#2}\left(#1\right)}

\def\xic{\xi_\mathrm{c}}

\def\fAl{f_\infty}
\def\fbig{\bar{f}}
\def\fA{\bar{f}_\infty}
\def\fAc{\bar{f}_\mathrm{c}}

\def\xic{\xi_{\mathrm{c}}}
\def\xim{\xi_{\mathrm{max}}}
\def\xin{\xi_{\mathrm{min}}}
\def\xia{[{\xi}]}

\def\B{\mathrm{T}}

\def\HB{H_{\B}}

\def\HS{H_\mathrm{D}}
\def\HI{V}

\def\WB{W_\mathrm{T}}
\def\hc{h_{\mathrm{c}}}

\def\cS{c_{\mathrm{S}}}
\def\cW{c_{\mathrm{W}}}

\newcommand{\specialcell}[2][c]{%
  \begin{tabular}[#1]{@{}c@{}}#2\end{tabular}}

\begin{document}
\title{Avalanche induced co-existing localised and thermal regions in disordered chains}
\author{P. J. D. Crowley}
\affiliation{Department of Physics, Boston University, MA 02215}
\author{A. Chandran}
\affiliation{Department of Physics, Boston University, MA 02215}

\date{\today}

\begin{abstract}
We investigate the stability of an Anderson localized chain to the inclusion of a single finite interacting thermal seed. 
This system models the effects of rare low-disorder regions on many-body localized chains.
Above a threshold value of the mean localization length, the seed causes runaway thermalization in which a finite fraction of the orbitals are absorbed into a thermal bubble. 
This `partially avalanched' regime provides a simple example of a delocalized, non-ergodic dynamical phase.
We derive the hierarchy of length scales necessary for typical samples to exhibit the avalanche instability, and show that the required seed size diverges at the avalanche threshold. 
We introduce a new dimensionless statistic that measures the effective size of the thermal bubble, and use it to numerically confirm the predictions of avalanche theory in the Anderson chain at infinite temperature.
\end{abstract}

\maketitle

\section{Introduction}

Sufficiently strong disorder localizes non-interacting particles in any dimension~\cite{anderson1958absence}.
Each single-particle orbital is then exponentially localized with a (energy-dependent) localization length.
Rare low disorder regions have no deleterious effects on such systems as the asymptotic decay of the associated orbitals is set by the disorder strength of the typical regions.

In one dimension $d=1$, finite but weak interactions between the particles preserve localization~\cite{basko2006metal,basko2006problem,oganesyan2007localization,vznidarivc2008many,bardarson2012unbounded,iyer2013many,de2013ergodicity,huse2013localization,kjall2014many,nandkishore2015many,lev2015absence,gopalakrishnan2015low,serbyn2015criterion,vasseur2015quantum,lim2016many,imbrie2016many,imbrie2016diagonalization,dumitrescu2017scaling,de2017quantum,abanin2017recent,pekker2017encoding,yu2017finding,alet2018many,abanin2019colloquium}.
In such `many-body localized' (MBL) systems, the orbitals are dressed into quasi-local integrals of motion called localized-bits (or l-bits)~\cite{serbyn2013local,serbyn2013universal,huse2014phenomenology,chandran2015constructing,ros2015integrals,pekker2017fixed,pancotti2018almost,varma2019length}.
These l-bits underlie the persistent local memory observed in several quantum optical experiments~\cite{schreiber2015observation,wei2018exploring,kondov2015disorder,bordia2017probing,choi2016exploring,rubio2018probing,smith2016many,kucsko2018critical,bordia2017periodically,luschen2017signatures}.

What of the effects of rare regions in interacting localized systems? 
Interacting rare regions are much better baths than their non-interacting cousins as their densities of states are exponentially larger~\cite{de2017stability,gopalakrishnan2016griffiths,agarwal2015anomalous,schulz2019phenomenology}.
They seed thermal bubbles which can grow by thermalizing nearby l-bits. As the bubble grows, it becomes a more effective bath that can absorb more distant l-bits. If this process runs away, then the system undergoes a thermalization `avalanche'.

De Roeck and Huveneers (DRH) argued that avalanches destabilise MBL at any disorder strength in $d>1$, so that MBL is an unstable dynamical phase~\cite{de2017stability}. In $d=1$ however, the instability is present only if the l-bit localization length exceeds a threshold value. 
Avalanche theory thus predicts the existence of stable MBL and thermalizing phases, and the properties of the phase transition between them~\cite{thiery2018many}.

Despite the far-reaching implications of the avalanche instability, it has been tested in relatively few studies. Refs.~\cite{luitz2017small,potirniche2018stability} numerically found the instability in toy models defined in the l-bit basis, while Ref.~\cite{ponte2017thermal} showed that a central spin can thermalize many l-bits at a vanishing interaction scale. However, a recent numerical study reports no evidence of avalanches in MBL spin chains~\cite{goihl2019exploration}. Furthermore, experiments with ultra-cold atoms suggest a surprising robustness of the MBL phase in two dimensions~\cite{rubio2018probing}.
Using two-component mixtures in an optical lattice with only one of the components experiencing a disorder potential, Ref.~\cite{rubio2018probing} reported persistent local memory in the disordered component despite interactions with the clean component.

\begin{figure}
    \centering
    \includegraphics[width=0.95\columnwidth]{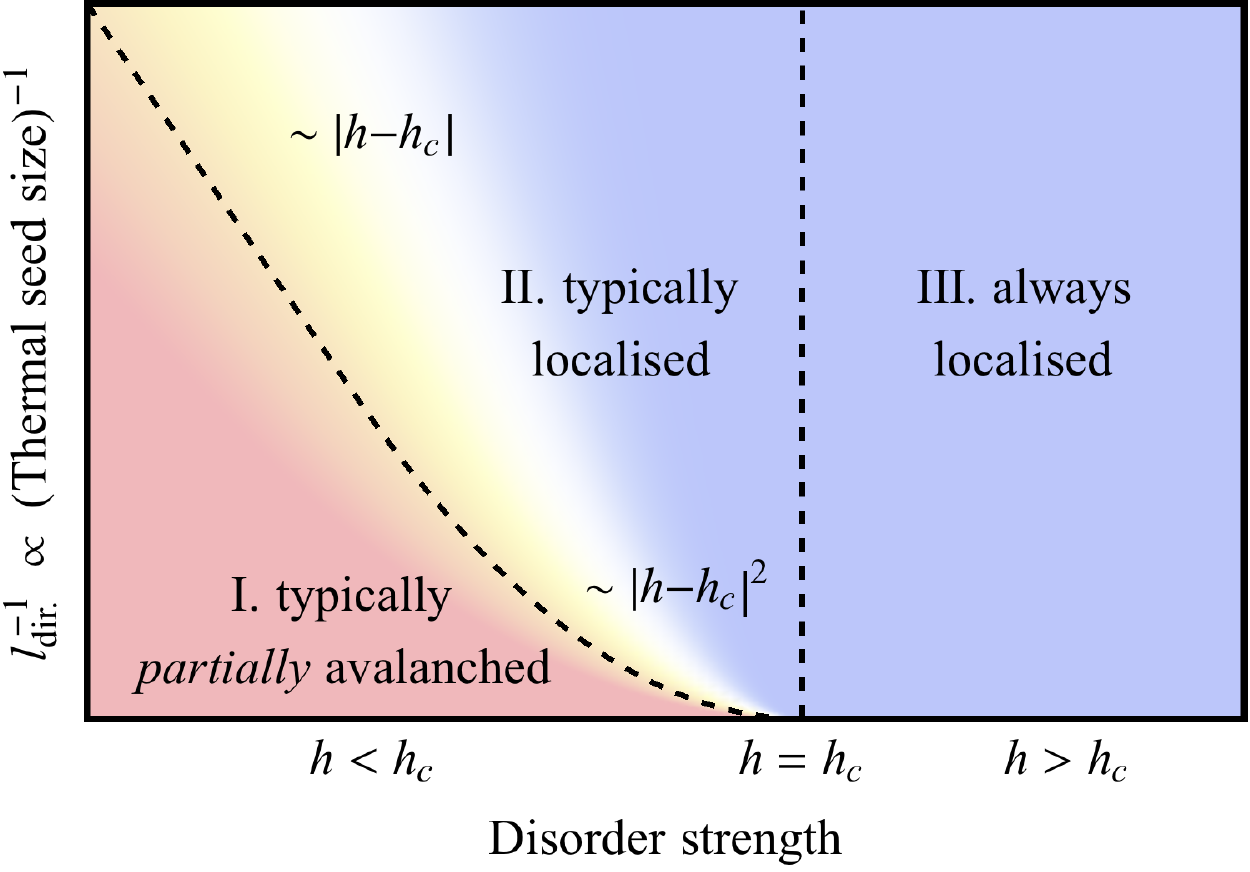}
    \caption{\emph{Phase diagram of avalanches}: Density plot illustrating the fraction of samples which avalanche close to the threshold. Typical samples avalanche (region I) only when $h < \hc$ and the initial thermal seed is sufficiently large, or more precisely, when the hierarchy of scales Eq.~\eqref{eq:hierarchy} is observed. As $h$ approaches $\hc$ from below, the necessary seed size for typical samples to avalanche diverges as $|h-\hc|^{-2}$ after pre-asymptotic scaling of $|h-\hc|^{-1}$. For $h < \hc$ with small seeds, rare samples may still avalanche (region II). For $h > \hc$ no samples avalanche (region III).
    }
    \label{Fig:CriticalFan}

\end{figure}

\begin{figure}
    \centering
    \includegraphics[width=0.95\columnwidth]{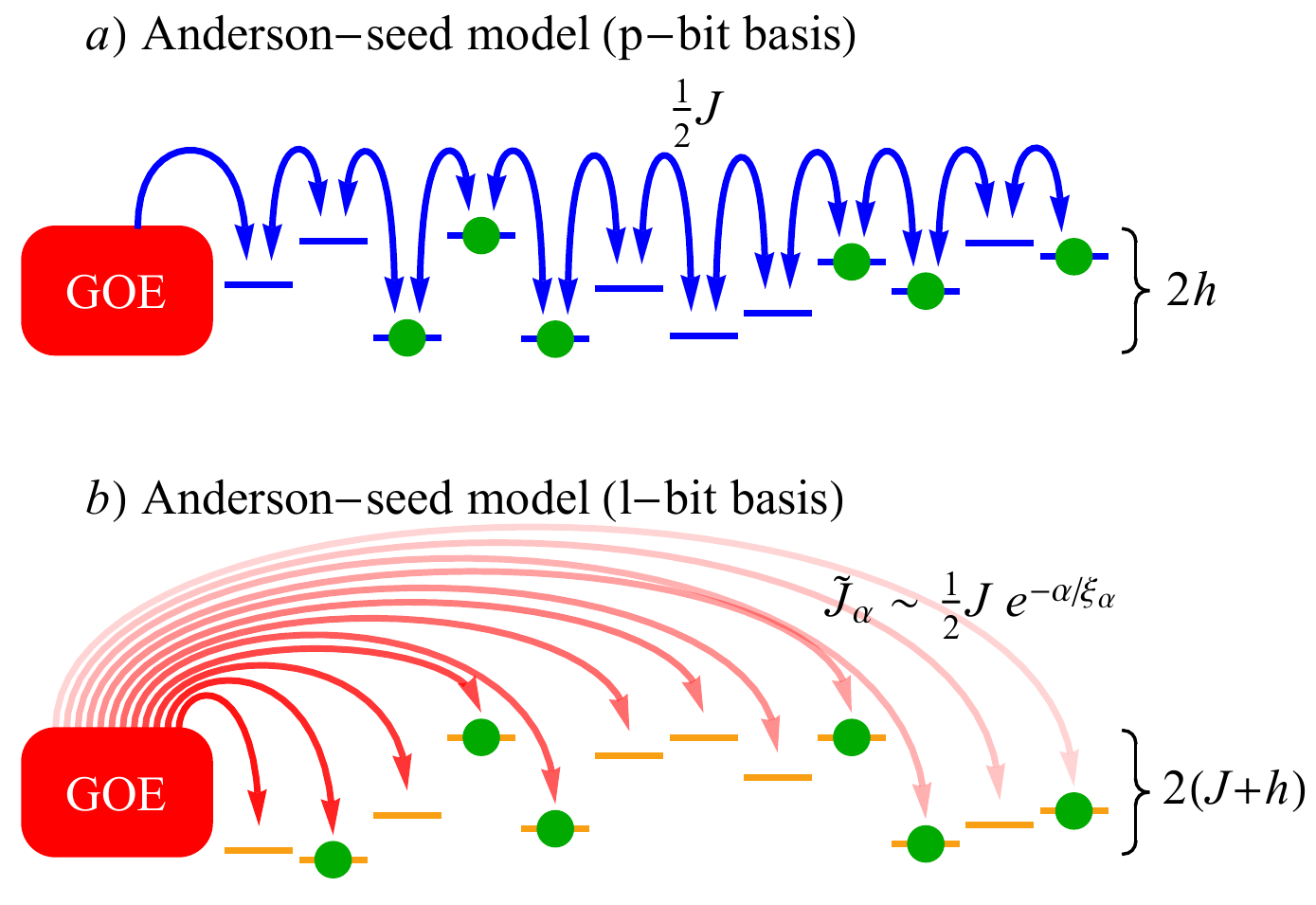}
    \caption{\emph{The Anderson-seed model:} An Anderson chain coupled to a thermal seed in (a) the physical (p-bit) basis, and (b) the diagonal (l-bit) basis. Under the Jordan Wigner transformation, the random-XX chain coupled to a thermal seed maps to this model. 
    }
    \label{fig:Cartoon}

\end{figure}

In this manuscript, we test the avalanche instability for an Anderson chain coupled to a single interacting thermal seed (Fig.~\ref{fig:Cartoon}).
We focus on the Anderson chain as the l-bits are uniquely defined and numerically accessible. The Anderson l-bits have a distribution of localization lengths, whereas DRH assumed the MBL l-bits to be characterised by a unique localization length~\footnote{We note that Ref.~\cite{varma2019length} provides numerical evidence that the localization length is unique deep in the MBL phase.}.

Our first result is that introducing a single thermal seed into a system with a distribution of localization lengths will result in a \emph{partial avalanche}. This occurs when the \emph{mean} localization length exceeds the critical value
\begin{align}
\xia > \xi_c \equiv \frac{2}{\log2}.
\label{Eq:MeanXiCondn}
\end{align}
When the system avalanches the resulting thermal bubble absorbs a non-contiguous finite fraction $\fA$ of the chain's l-bits, and co-exists with the remaining localized l-bits. The partially avalanched Anderson-seed model thus provides an example of a delocalized non-ergodic phase \red{that is robust to all perturbations which preserve the free-fermion structure of the chain Hamiltonian.} For the Anderson chain with box disorder, we compute $\fA$ as a function of the disorder strength $h$ (Fig.~\ref{fig:PAval}). 

Next, for $\xia > \xic$ (or equivalently for disorder strengths $h < \hc$), we identify the length scale up to which the thermal bubble must grow before the instability argument ensures that typical samples exhibit avalanches. As this length scale diverges as $h \to \hc$ from below (with exponent $\nu = 2$), the probability that a thermal seed of fixed finite size sets up a runaway avalanche vanishes in the same limit. Fig.~\ref{Fig:CriticalFan} summarises the behaviour of the ensemble averaged fraction of l-bits absorbed into the thermal bubble $\fbig$ as a function of $h$ and the thermal seed size. 

In more detail, we identify three length scales: $l_{\mathrm{dir}}$, $l_\mathrm{FS}$ and $l^\star$. The scale $l_{\mathrm{dir}}$ sets the number of l-bits with significant direct coupling to the bare thermal seed, and is proportional to the number of degrees of freedom in the thermal seed (Eq.~\eqref{Eq:ldirRelation}). The second scale $l_\mathrm{FS}$ is the Harris-Luck scale, this sets the number of l-bits required to accurately determine whether the disorder strength $h$ is above or below the threshold value $\hc$~\cite{harris1974effect,chayes1986finite,luck1993classification,chandran2015finite}. Thermal bubbles of spatial extent greater than the third scale $l^\star$ typically grow indefinitely if $h$ can be determined accurately on the scale $l^\star$, i.e. if $l^\star \gg l_\mathrm{FS}$. 

We present analytical arguments that typical samples avalanche if:
\begin{equation}
    L \gg l_{\mathrm{dir}} \gg l^\star, l_{\mathrm{FS}}.
    \label{eq:hierarchy}
\end{equation}
As $l^\star$ and $l_\mathrm{FS}$ diverge near $h=\hc$,
\begin{equation}
\begin{aligned}
l^\star &\sim |h - \hc|^{-1}, \\
l_{\mathrm{FS}} &\sim |h - \hc|^{-2},
\end{aligned}
\end{equation}
while the seed size and thus $l_\mathrm{dir}$ remain finite, typical samples partially avalanche only in region I of Fig.~\ref{Fig:CriticalFan}. In contrast, in region II, though rare samples may avalanche, in typical samples the thermal bubble is of finite extent.

Third, we derive the conditions on the coupling strength between a single l-bit and a thermal seed which determine when the two systems are either very strongly or very weakly hybridised (Eqs.~(\ref{eq:condition_coupling},\ref{eq:weak_coupling})). On applying these conditions to the Anderson-seed model near the avalanche threshold, we obtain the remarkable result that there are approximately \emph{ten l-bits with intermediate values of couplings} to the bare thermal seed. This provides conservative bounds on $l_\mathrm{dir}$: 
\begin{align}
l_{\mathrm{S}} \, \leq \, l_{\mathrm{dir}} \, \leq \, l_{\mathrm{S}} + 9.9,
\label{Eq:ldirRelation}
\end{align}
where $l_\mathrm{S}$ is proportional to the number of degrees of freedom in the thermal seed. This implies that most l-bits in numerically accessible finite-size chains are in the intermediate regime in which the extent of hybridisation between the l-bit and the seed is partial with large eigenstate-to-eigenstate variations (Appendix~\ref{app:cScW}).

As we find significant seed sizes to be necessary, the chain lengths $L$ accessible to exact diagonalization are $L \approx 10 \sim l_{\mathrm{dir}}$. As Eq.~\eqref{eq:hierarchy} is not satisfied, it is difficult to conclusively identify avalanches. Furthermore, the partially avalanched Anderson-seed system is delocalized, but non-ergodic. These phases are also notoriously hard to characterise numerically~\cite{de2014anderson,altshuler2016nonergodic,kravtsov2018non,biroli2018delocalization}.

To address the challenge of identifying the non-ergodic delocalized phase, we present our fourth result: a new dimensionless statistic $[v]$ which measures the size of a non-contiguous thermal bubble. This statistic characterises the seed connected to the Anderson chain by its effectiveness as a bath for a probe l-bit. Specifically, we use a modified ratio of the matrix elements between eigenstates to their energy level spacing to determine the number of degrees of freedom in the total system that couples to the probe l-bit \red{(previously introduced in Ref.~\cite{thiery2018many})}. The measure $[v]$ grows linearly with $L$ with slope $\fbig/\xic$ if and only if a finite fraction $\fbig$ of the l-bits in the chain have been absorbed into a thermal bubble. We see robust evidence for linear growth for $\xia > \xi_c$ in the Anderson-seed system. 

As an aside, we also confirm that $[v]$ detects the MBL transition in the interacting disordered fermionic chain (corresponding to the disordered Heisenberg model). \red{We find that the finite-size estimate of the critical disorder strength agrees with that extracted from standard spectral measures, despite the inclusion of an external thermal seed (Sec.~\ref{sec:numerics}).}

The outline of the paper is as follows. In Sec.~\ref{sec:model} we introduce the model of interest, a disordered fermionic/spin chain coupled to a thermal seed. In Sec.~\ref{sec:starting}, we derive the conditions under which the thermal seed absorbs the first few l-bits. We subsequently generalise the theory of DRH in Sec.~\ref{sec:runaway_theory}, and show that due to fluctuations on the localization length the avalanche instability only leads to at most partial thermalization. We then turn to the length scales that control the probability that a finite thermal seed can set up a runaway (partial) avalanche in Sec.~\ref{sec:threshold}. After deriving further conditions on the Anderson-seed model to see the runaway avalanche instability in Sec.~\ref{sec:runaway_conditions}, we turn to finite-size numerics in Sec.~\ref{sec:numerics}. We define the new dimensionless measure $[v]$ and present numerical evidence in favour of partial avalanches in the Anderson-seed model. Finally, we discuss the implications of this work for MBL systems in Sec.~\ref{sec:conclusions}.

\section{Model}
\label{sec:model}

The random-XXZ chain provides a canonical model of both single-particle and many-body localization:
\begin{equation}
\HS =  J\sum_{n=1}^{L-1} \left( s_n^x s_{n+1}^x + s_n^y s_{n+1}^y + \Delta s_n^z s_{n+1}^z \right) + \sum_{n=1}^L h_n s_n^z.
\end{equation}
Here $s_n^\nu = \tfrac12 \sigma_n^\nu$ are the usual spin-$\tfrac12$ operators, the local fields are drawn independently from the box distribution $h_n \in [-h,h]$, and we use open boundary conditions.
Under the Jordan-Wigner transformation, the XXZ chain maps to an interacting disordered fermionic model~\cite{jordan1928pauli,lieb1961two}:
\begin{equation}
\begin{aligned}
    \HS = & \frac{J}2 \sum_{n=1}^{L-1} \left( a_n^\dag a_{n+1} + \mathrm{h.c.} \right) + \sum_n \mu_n a_n^\dag a_{n} 
    \\ & \quad \quad + J \Delta \sum_{n=1}^{L-1} a_n^\dag a_{n} a_{n+1}^\dag a_{n+1} 
    \label{eq:Ham_ferm}
\end{aligned}
\end{equation}
up to an additive constant. Above,
\begin{equation}
\mu_n =
    \begin{cases}
    h_n - J \Delta & \text{ for } 1 < n < L
    \\
    h_n - \tfrac12 J \Delta & \text{ for } n = 1,L
    \end{cases}
\end{equation}
and the fermionic creation and annihilation operators satisfy the usual anti-commutation relations $\{a_n, a_m^\dagger\}=\delta_{mn}$.

In this article, we focus on two values of $\Delta$ in the random-XXZ chain.
\begin{enumerate}
\item The random-XX chain ($\Delta=0$), which maps to the non-interacting Anderson model Eq.~\eqref{eq:Ham_ferm}~\cite{anderson1958absence}.
\item The random-Heisenberg chain ($\Delta=1$), which maps to an interacting and disordered fermionic model. 
\end{enumerate}

\red{
To probe the avalanche instability, we couple one end of the disordered chain to an external thermal seed with even Hilbert space dimension $d_0$. The total Hamiltonian acts on the Hilbert space obtained by taking the tensor product of the Hilbert spaces of the thermal seed and the chain and is given by:
\begin{equation}
    H = \HB \otimes \mathbf{1} + \mathbf{1} \otimes \HS + \HI.
    \label{eq:fullH}
\end{equation}
The Hamiltonian is depicted in Fig~\ref{fig:Cartoon}(a) in the fermionic representation. We discuss the first and third term in Eq.~\eqref{eq:fullH} in turn below.
}

\red{
The Hamiltonian of the thermal seed $\HB$ is given by a $d_0 \times d_0$ random matrix of bandwidth $4 \WB$ drawn from the Gaussian Orthogonal Ensemble (GOE). 
\red{We take $d_0$ to be an even integer; we explain why below.}
The eigenvalues $w_a$ of $\HB$ have mean $[w_a]=0$ and variance $[w_a^2] = \WB^2$.
The square brackets $[\cdot]$ denotes ensemble averaging with respect to the GOE matrix $\HB$ and the on-site disorder in the spin chain. We refer to a single instance of this ensemble as a sample.
An important quantity in subsequent analysis is the density of states of the thermal seed at maximal entropy:
\begin{align}
\rho_0 = \frac{d_0}{\pi \WB}.
\end{align}
}

\red{For even $d_0$, we define spin-$\tfrac12$ operators on the thermal seed as follows. Let $d_0 \in 2 \mathbb{N}$. Such a thermal seed may be obtained by coupling a $\frac{d_0}{2}$-level system to a spin-$\tfrac12$ particle (with label $\B$). The required seed spin-$\tfrac12$ operators are then $s_\B^{\nu}$ for $\nu=x,y,z$. Furthermore, the Jordan-Wigner transformation on the spins with labels $\{\B, 1, 2, \ldots L\}$ defines a fermionic creation operator $a_{\B}^\dag$ on the thermal seed that satisfies the usual anti-commutation relations $\{a_n, a_m^\dagger\}=\delta_{mn}$ for $n,m \in \{\B,1,2, \ldots L\}$.}

\red{The decomposition of the thermal seed into a $\frac{d_0}{2}$-level system and spin-$\tfrac12$ particle is useful to define the seed-chain interaction term $V$ in Eq.~\eqref{eq:fullH} to have the same form as the term in $\HS$ that couples neighbouring sites:}
\begin{equation}
\begin{aligned}
    \HI & =  J \Big( s_{\B}^x \otimes s_{1}^x + s_{\B}^y  \otimes s_{1}^y + \Delta s_{\B}^z \otimes s_{1}^z \Big),
    \\
    & = 
    \frac{J}{2}
    \begin{pmatrix} 
    a_{\B}^\dag  &
    a_{1}^\dag  
    \end{pmatrix}
    \begin{pmatrix} 
    - \Delta & 1 \\
    1 & - \Delta
    \end{pmatrix}
    \begin{pmatrix} 
    a_{\B}  \\
    a_{1}  
    \end{pmatrix}
    + \Delta J a_{\B}^\dag a_{\B} a_{1}^\dag a_{1}.
\end{aligned}
\end{equation}
See Fig~\ref{fig:Cartoon}(a).

\subsection{l-bit basis}

In the Anderson model ($\Delta = 0$), $\HS$ is bi-linear in the fermionic operators, and hence easily diagonalized
\begin{equation}
    \begin{aligned}
        \HS & = \sum_{\alpha=1}^L {\tilde\epsilon}_\alpha f_\alpha^\dag f_\alpha,    
        \label{Eq:AndersonLbitHam}
    \end{aligned}
\end{equation}
where $f_\alpha^\dag = \sum_n \phi_n^\alpha a_n^\dag$ creates a fermion in the $\alpha$th diagonal orbital, and the single particle energies ${\tilde\epsilon}_\alpha$ and orbitals $\phi_\alpha^n$ are the solutions to the equation:
\begin{equation}
    J \phi^\alpha_{n+1} + J \phi^\alpha_{n-1} + 2 h_n \phi^\alpha_{n} = 2 {\tilde\epsilon}_\alpha \phi^\alpha_{n}.
\end{equation}
The diagonal orbitals define the localized-bits (or l-bits) of the disordered chain in the absence of coupling to the seed ($V=0$)~\cite{serbyn2013local,serbyn2013universal,huse2014phenomenology,chandran2015constructing,pekker2017fixed}. 

The single particle energies have mean and variance given by:
\begin{equation}
    \begin{aligned}
        {[{\tilde\epsilon}_\alpha]} & = 0,  \quad & {[{\tilde\epsilon}_\alpha^2]} & = \tfrac12 J^2 + \tfrac13 h^2.
    \end{aligned}
\end{equation}
They are further bounded by the following inequality (which is saturated as $L \to \infty$) 
\begin{equation}
    - J - h \, \leq \, {\tilde\epsilon}_\alpha \, \leq \, J + h 
    \label{Eq:SingleParticleEnergyBound}
\end{equation}
The l-bit orbitals $\phi_n^\alpha$ are exponentially localized~\cite{anderson1958absence} with localization length $\xi_\alpha$, and localization centres
\begin{equation}
    \bar{n}_\alpha = \sum_n n |\phi_n^\alpha|^2.
\end{equation}
We index the orbitals according to their distance from the thermal seed so that $\alpha < \beta \Longleftrightarrow \bar{n}_\alpha < \bar{n}_\beta$. This indexing scheme implies that $[|\bar{n}_\alpha - \alpha|] = \mathrm{O}(\xi_\alpha)$. 

Re-writing the seed-chain coupling $V$ in the diagonal basis:
\begin{align}
\HI & =  \sum_{\alpha=1}^L {\tilde{J}}_\alpha \big( a_{\B}^\dag \otimes f_\alpha + \mathrm{h.c.} \big).
\end{align}
The $\alpha$th l-bit is coupled to the seed with strength that decays exponentially with $\alpha$:
\begin{equation}
    \tilde{J}_\alpha = \tfrac12 J \phi_1^\alpha \sim J \e^{-\alpha/\xi_\alpha}
    \label{eq:couplings}
\end{equation}
The Hamiltonian of the Anderson-seed system in the l-bit basis is shown in Fig.~\ref{fig:Cartoon}(b). 

\red{The interacting fermion system with $\Delta \neq 0$ is more complex} as $\HS$ cannot be simply diagonalized.
Nevertheless, full localization persists above a non-zero critical disorder strength~\cite{basko2006metal,basko2006problem,oganesyan2007localization,vznidarivc2008many,bardarson2012unbounded,iyer2013many,de2013ergodicity,huse2013localization,kjall2014many,nandkishore2015many,lev2015absence,gopalakrishnan2015low,serbyn2015criterion,vasseur2015quantum,lim2016many,imbrie2016many,imbrie2016diagonalization,dumitrescu2017scaling,de2017quantum,abanin2017recent,pekker2017encoding,yu2017finding,alet2018many,abanin2019colloquium}.
This many-body localized (MBL) phase is characterised by an extensive set of quasi-local integrals of motion (the l-bits) which generalise the diagonal orbitals of the Anderson model to the interacting case~\cite{serbyn2013local,serbyn2013universal,huse2014phenomenology,chandran2015constructing,ros2015integrals,pekker2017fixed,pancotti2018almost,varma2019length}.
The expansion of $H$ in terms of the l-bit operators contains higher-order terms as compared to the Anderson case:
\begin{equation}
    \begin{aligned}
        \HS = &  \sum_{\alpha} {\tilde\epsilon}_\alpha f_\alpha^\dag f_\alpha + \sum_{\alpha \beta} {\tilde K}_{\alpha\beta}^{(2)}  f_\alpha^\dag f_\alpha f_\beta^\dag f_\beta + \ldots
        \\
        \HI = & a_{\B}^\dag \otimes \left(  \sum_{\alpha} {\tilde J}_\alpha^{(1)} f_\alpha + \sum_{\alpha\beta\gamma} {\tilde J}_{\alpha\beta\gamma}^{(3)} f_\alpha^\dag f_\beta f_\gamma + \ldots \right) + \text{h.c}.
    \end{aligned}
    \label{eq:VV}
\end{equation}
The higher-order terms in $\HS$ encode diagonal interaction energies between the l-bits. These interaction energies decay exponentially with the distance between the l-bits~\cite{nandkishore2015many,pekker2017fixed,villalonga2018exploring,varma2019length}.
We ignore these diagonal interaction energies henceforth.
The higher order terms in $V$ allow several l-bits to reconfigure simultaneously through their interaction with the thermal seed. 
\red{The typical strength of interaction terms that reconfigure the $\alpha$th l-bit (i.e. terms ${\tilde J}_\alpha^{(n)}$ in $V$ which do not commute with the l-bit occupation number $f_\alpha^\dag f_\alpha$) decay exponentially in the order $n$, however as there are exponentially many-in-$n$ such terms they cannot be discarded by any naive argument.  \redd{From the Heisenberg equation of motion we see that the energy scale associated with l-bit flip processes is given by
\begin{equation}
    \left| \frac{\d f_\alpha^\dag f_\alpha}{\d t} \right| = \left|[H,f_\alpha^\dag f_\alpha] \right| = \left|[V,f_\alpha^\dag f_\alpha] \right|.
\end{equation}
Thus the effective coupling $\tilde{J}_\alpha$ between the $\alpha$-th l-bit is given by the norm of the commutator:}} \redd{
\begin{equation}
    \tilde{J}_\alpha : = |[V,f_\alpha^\dag f_\alpha]| \sim J \e^{- \alpha/\xi_\alpha}.
    \label{eq:Vcomm}
\end{equation}
The coupling $\tilde{J}_\alpha$ determines the interaction energy scale and may be understood as a re-summation of the terms ${\tilde J}_\alpha^{(n)}$.} Thus, the qualitative features of the Hamiltonian in the l-bit basis in the interacting case is the same as in Fig.~\ref{fig:Cartoon}(b) with $\xi_\alpha$ in Eq.~\eqref{eq:Vcomm} defining the effective localization length of the $\alpha$-th l-bit.

\red{We note that the effective localization length defined through Eq.~\eqref{eq:Vcomm} may be longer than that controlling the decay of any single bare coupling, e.g.  $\tilde{J}_\alpha^{(1)}$ in Eq.~\ref{eq:VV}, as multiple interaction terms contribute to the decay. This discrepancy has been observed numerically in Ref.~\cite{varma2019length}}.

\subsection{The localization length distribution in Anderson chains}
The envelope of each single particle orbital decays exponentially
\begin{align}
 |\phi_n^\alpha| \sim \e^{-|n - \alpha|/\xi_\alpha}
 \label{Eq:PhiDecay}
 \end{align}
 with a characteristic localization length $\xi_\alpha$. This localization length varies as a smooth function of the single particle energy as $L \to \infty$,
\begin{align}
\xi_\alpha = \xi({\tilde\epsilon}_\alpha).
\end{align}
Thus, the distribution of localization lengths
\begin{equation}
    p(\xi') : = \frac{1}{L} \sum_\alpha \delta(\xi' - \xi_\alpha )
\end{equation}
converges to a simple analytic form set by the single particle density of states $g(\epsilon)$
\begin{equation}
\lim_{L \to \infty} p(\xi') = \int d\epsilon \, g(\epsilon) \, \delta(\xi'-\xi(\epsilon)).
\end{equation}
The distribution $p(\xi)$ is easily interpreted: consider the orbitals $1 < \alpha < l$ which are centred within a region of length $l \gg 1$ of the Anderson chain. The expected number of orbitals in this region with localization lengths $\xi_\alpha \in [\xi,\xi+d \xi]$ is given by $l p(\xi) d \xi$. 

\section{Starting an avalanche: absorbing the first few l-bits}
\label{sec:starting}

In Ref.~\cite{de2017stability}, DRH studied the effects of coupling a finite thermal seed to a localized chain.
They assumed that the \emph{avalanche could start}, that is, that the seed hybridised strongly with nearby l-bits and forms a small \emph{thermal bubble}.
They then derived conditions that the thermal bubble grows asymptotically, so that l-bits that are very distant from the seed eventually hybridise with the thermal bubble and reach thermal equilibrium.
If the avalanche does not start, then the asymptotic thermalization instability is immaterial.

Sec.~\ref{sec:threshold} discusses several length scales that control the probability to start an avalanche and their requisite hierarchy for a typical sample to avalanche.
In this section, we have a more modest aim: we identify parameter regimes of $H$ in which the bare thermal seed strongly hybridises with a \emph{non-zero} number of nearby l-bits  $l_\mathrm{S}$ due to the direct interactions in $ \HI$:
\begin{equation}
    l_\mathrm{S} > 0.
\end{equation}
The above condition is necessary, but not sufficient for starting an avalanche with high probability. We return to the sufficient conditions in Sec.~\ref{sec:runaway_conditions}.

\subsection{Bandwidth Condition: There exist resonances between the thermal seed and nearby l-bits}

The first l-bit can flip only if it can exchange an amount of energy ${\tilde\epsilon}_1$ with the thermal seed. For small $J$, we thus require the seed bandwidth $4\WB$ to be larger than ${\tilde\epsilon}_1$. Using the bound on single-particle energies in Eq.~\eqref{Eq:SingleParticleEnergyBound}, we obtain the condition in Eq.~\eqref{eq:condition_bandwidth} for the Anderson model.  

When $\Delta \neq 0$, and indeed in general MBL models, there is no known general bound on $|{\tilde\epsilon}_\alpha|$. 
Nevertheless, at strong disorder strengths, we expect that, up to small corrections, $\HS \approx \sum_n \tilde{\epsilon}_\alpha f_\alpha^\dagger f_\alpha $ and $\tilde{\epsilon}_\alpha \approx h_\alpha$.
The first relationship implies that the variance of the l-bit energies is set by the variance of $\HS$:
\begin{align}
[\tilde{\epsilon}_\alpha^2] \approx 4 \frac{\tr{\HS^2}-(\tr\HS)^2}{L 2^L}.
\end{align}
Using a property of the box distribution that $|h_\alpha| \leq \sqrt{3[h_\alpha^2]}$, we obtain the following relation at strong disorder
\begin{equation}
\begin{aligned}
    |\tilde{\epsilon}_\alpha| \approx  |h_\alpha| \leq  \sqrt{3[h_\alpha^2]} &\approx \sqrt{3[{\tilde\epsilon}_\alpha^2]} \\
   & \lesssim \sqrt{\tfrac32 J^2+ \tfrac34\Delta^2 J^2 + h^2}.
\end{aligned}    
\end{equation}

In summary, enforcing the condition that $4 \WB > |{\tilde\epsilon}_1|$ gives:
\begin{equation}
    4 \WB > \begin{cases}
    J + h & \qquad \text{(Anderson)},
    \\
    \sqrt{\tfrac32 J^2+ \tfrac34\Delta^2 J^2 + h^2} & \qquad \text{(Heisenberg)}.
    \end{cases}
    \label{eq:condition_bandwidth}
\end{equation}
In numerical experiments, this bound must be satisfied by a reasonable margin as the density of states of the thermal seed vanishes at the edge of its spectrum. 

\subsection{Coupling strength condition: The thermal seed is sufficiently strongly coupled to nearby l-bits}
\label{sec:goodthermalseed}

For $\HI=0$ the eigenstates of $H$ are product states of the thermal seed and a configuration of the l-bits in the disordered chain $\ket{\psi_i} = \ket{w_a} \otimes \ket{c_b}$. \redd{Here $i = (a,b)$ where $a = 1 \ldots d$ enumerates the states of the bath and $b = 1 \ldots 2^L$ enumerates the states of the chain.} Consider the fate of the product eigenstate $\ket{\psi_i} = \ket{w_a} \otimes \ket{c_b}$ when interactions are reintroduced.
Let the closest l-bit with index $\alpha=1$ be unoccupied, i.e. $f_1 \ket{c_b} = 0$.
At small $\HI$, the l-bit hybridises with the seed if the typical matrix element $\HI_{ij}$ between $\ket{\psi_i}$ and a second eigenstate $\ket{\psi_j} = \ket{w_c} \otimes \ket{c_d}$ exceeds the typical energy level spacing of the seed \redd{$\delta_{ij} \sim \rho_0^{-1}$. Here $\delta_{ij} = |E_i - E_j|$ is the separation of the energies of the state $\ket{\psi_i}$, $\ket{\psi_j}$.}
The typical matrix element is given by:
\begin{equation}
\begin{aligned}
    V_{ij} = \bra{\psi_i}\HI\ket{\psi_j} &= \tfrac12 J \phi_1^\alpha \cdot \bra{w_a} a_{\B}^\dag \ket{w_c} \cdot \bra{c_b} f_1 \ket{c_d} 
    \\
    &\sim {\tilde J}_\alpha \cdot \frac{1}{\sqrt{d_0}} \cdot 1.
\end{aligned}
\label{eq:RMT1}
\end{equation}
Hybridisation requires that typically $V_{ij} \gg \rho_0^{-1}$. \redd{In the Heisenberg case the interaction in the l-bit basis is more highly structured. However the structure of matrix elements remains simple, specifically we take the matrix elements of local operators to follow eigenstate thermalization hypothesis (ETH)~\cite{srednicki1994chaos,deutsch1991quantum,dalessio2016quantum}. From ETH it follows that the distribution of matrix elements is characterised by a single scale, here ${\tilde J}_\alpha$, and Eq.~\eqref{eq:RMT1} follows.}

How much larger should $V_{ij}$ be as compared to $\rho_0^{-1}$ in order for the l-bit to be strongly hybridised? This is tricky to quantify as the distribution of \redd{$|V_{ij}/\delta_{ij}|$} is broad even within a sample. In Appendix~\ref{app:cScW}, we use a second probe l-bit to define a statistic $[v]$ (detailed in Sec.~\ref{sec:numerics}) which depends on the \redd{$|V_{ij}/\delta_{ij}|$}. Using $[v]$ we determine that a thermal seed hybridises strongly with a single l-bit if
\begin{equation}
    \frac{{\tilde J}_\alpha \rho_0}{\sqrt{d_0}} > \cS \approx 0.31
    \label{eq:condition_coupling}
\end{equation}
When Eq.~\eqref{eq:condition_coupling} is satisfied by the first $l_\mathrm{S}$ l-bits, the seed strongly hybridises with these l-bits and the thermal bubble grows by $l_\mathrm{S}$ l-bits.

Similarly, the typical configurations of the thermal seed and l-bit do not hybridise significantly if
\begin{equation}
    \frac{{\tilde J}_\alpha \rho_0}{\sqrt{d_0}} < \cW \approx 0.01.
    \label{eq:weak_coupling}
\end{equation}
When Eq.~\eqref{eq:weak_coupling} is satisfied by the first l-bit, the seed and l-bit remain very weakly coupled. Consequently, the thermal bubble does not grow larger than the seed and the avalanche does not start.

We observe that $\cW$ and $\cS$ are approximately two orders of magnitude apart. At values of $\tilde{J}_\alpha \rho_0/\sqrt{d_0}$ between $\cW$ and $\cS$, the l-bit is partially absorbed into the thermal bubble and there is significant eigenstate-to-eigenstate variation in the degree of absorption. As the numerically accessible chain lengths are short, most l-bits are in this intermediate coupling regime in our numerical study in Sec.~\ref{sec:numerics}.

\section{Theory of the runaway instability}
\label{sec:runaway_theory}

We first review the avalanche theory of DRH~\cite{de2017stability} and derive the condition for the runaway thermalization instability for a unique l-bit localization length.
We then extend the avalanche theory to the case of finite localization length distributions and apply it to the Anderson-seed model.
We identify the critical disorder strength $\hc$ below which the instability is present, and discuss the properties of the partially avalanched regime.

In order to differentiate between complete and partial avalanches, we define $\fbig \in [0,1]$ to be the fraction of l-bits that are a part of the thermal bubble when the avalanche ceases.
In \emph{infinitely long chains}, the fraction $\fbig$ depends on the disorder strength $h$ and the number of l-bits that are directly thermalized by the seed $l_{\mathrm{dir}}$.

In this section, we assume that the avalanches start with probability one.
That is, we study the functional dependence of $\fbig$ on $h$ in the following limit:
\begin{align}
\fA( h) := \lim_{l_{\mathrm{dir}} \to \infty} \fbig(h, l_{\mathrm{dir}})
\end{align}
We emphasise the order of limits $L \to \infty$ first so that $l_{\mathrm{dir}} \ll L$ always.
DRH predict that $\fA(h)$ is a step function of unit height. Our extended avalanche theory predicts a non-trivial function $\fA( h)$ for $ h <\hc$ in the Anderson-seed model (horizontal axis of Fig.~\ref{Fig:CriticalFan}), with an intermediate critical value
\begin{equation}
    \fAc : = \lim_{h \to \hc^-}\fA( h)
\end{equation}

\subsection{Review of DRH Avalanche theory}
\label{Sec:DRHTheory}
The set-up in Ref.~\cite{de2017stability} is identical to that in Fig.~\ref{fig:Cartoon}(b) with all $\xi_\alpha$ equal. Suppose the starting seed has strongly hybridised with the first $(\alpha-1)$ l-bits with:
\begin{align}
\alpha \gg l_{\mathrm{dir}}.
\end{align} 
The thermal bubble then has Hilbert space dimension $d = d_0 2^{\alpha-1}$ and density of states $\rho = \rho_0 2^{\alpha-1}$. That is, the thermal bubble has an exponentially larger density of states as compared to the bare seed. From Eq.~\eqref{eq:condition_coupling}, the l-bit with index $\alpha$ strongly hybridises with this thermal bubble if its direct coupling is large enough:
\begin{align}
 \frac{{\tilde J}_\alpha \rho}{\sqrt{d}} = {\tilde J}_\alpha \rho_0 \sqrt{\frac{ 2^{\alpha-1}}{d_0}} > \cS
\end{align}
Using Eq.~\eqref{eq:couplings}, we find:
\begin{align}
\frac{J \rho_0}{\sqrt{2d_0}} \e^{-\alpha\left(\frac{1}{\xi_\alpha} - \frac{\log 2}{2}\right)} > \cS
\end{align} 
The above condition is satisfied by a greater and greater margin with increasing $\alpha$ provided that the localization length of each l-bit is above the threshold value
\begin{equation}
    \xi_\alpha > \xic = \frac{2}{\log 2}.
    \label{eq:naive}
\end{equation}
DRH thus predict complete thermalization \`{a} la ETH with $\fA=1$ when Eq.~\eqref{eq:naive} holds, and localization with $\fA=0$ otherwise. Numerical studies on toy models in the l-bit basis confirm this prediction~\cite{luitz2017small,potirniche2018stability}.

\subsubsection*{\red{Spectral function of the seed operator}}
\red{
The DRH avalanche theory assumes that the ETH holds in the thermal bubble at every stage of the avalanche process. Precisely, the theory requires that after $\alpha$ l-bits have been absorbed into the thermal bubble, the matrix elements of the seed operator are given by the ansatz
\begin{equation}
    \bra{b_a} a_{\B}^\dag \ket{b_b} = \sqrt{\frac{\mathcal{F}_{\mathrm{B}}(\omega) }{\rho}} \eta_{ab} \sim \frac{1}{\sqrt{d_0 2^{\alpha - 1}}}
    \label{Eq:ETHThermalBubble}
\end{equation}
where $\ket{b_a}$, $\ket{b_b}$ are eigenstates of the thermal bubble with respective energies $b_a$ and $b_b$ corresponding to infinite temperature, $\omega = b_a - b_b$, $\rho=\rho_0 2^{\alpha-1}$ is the relevant bubble density of states, $\mathcal{F}_{\mathrm{B}}(\cdot)$ is the spectral function of $a_{\B}^\dag$ in the thermal bubble, and $\eta_{ab}$ are random numbers with zero mean and unit variance. One might be concerned that this ansatz is inaccurate because the larger the $\alpha$, the (exponentially) longer it takes for the $\alpha$-th l-bit to be absorbed into the thermal bubble, and thus the smaller the scale of structure in the spectral function $\mathcal{F}_{\mathrm{B}}(\omega)$. Specifically, the $\alpha$th l-bit decays at a rate set by Fermi's Golden Rule
\begin{equation}
    \Gamma_\alpha = \left( \frac{{\tilde J}_\alpha}{\sqrt{d}} \right)^2 \rho \sim J^2 \rho_0 \e^{ - 2\alpha/\xi_\alpha}
\end{equation}
implying that the peaks of the spectral function have width $\Gamma_\alpha$. The ETH ansatz in Eq.~\eqref{Eq:ETHThermalBubble} is valid in a small energy window if this scale far exceeds level spacing, i.e. $\Gamma_\alpha \rho \gg 1$. As
\begin{equation}
    \Gamma_\alpha \rho = \left( \frac{{\tilde J}_\alpha \rho}{\sqrt{d}} \right)^2 \sim J^2 \rho^2_0\e^{2\alpha(1/\xic - 1/\xi_\alpha)},
\end{equation}
exponentially grows with $\alpha$ when $\xi_\alpha > \xic$, Eq.~\eqref{Eq:ETHThermalBubble} holds with increasing accuracy as the size of the thermal bubble increases in the avalanching regime. Refs.~\cite{potirniche2018stability} and~\cite{de2017stability} also discuss this `back-action' of the absorbed l-bits on the spectral function. As both studies conclude that more detailed treatments only provide sub-leading corrections to the DRH theory, we do not discuss the issue of back-action further.
}

\begin{table*}
  \renewcommand{\arraystretch}{1.45}
\begin{ruledtabular}
\begin{tabular}{ R{0.2\textwidth} || C{0.25\textwidth} | C{0.25\textwidth} | C{0.25\textwidth} }

& \specialcell{No Avalanche \\ (many-body localized)}
& \specialcell{Partial avalanche \\ (non-ergodic, delocalized)}
& \specialcell{Complete avalanche \\ (ergodic, delocalized)}
\\
\hline
Nearest neighbour energy level repulsion
& no
& no
& yes
\\
Scaling of eigenstate entanglement entropy
& area law  
& sub-thermal volume law with coefficient $<s_{\mathrm{th}}$
& volume law with coefficient $s_{\mathrm{th}}$
\\ 
Eigenstate entanglement entropy of individual p-bits
& $\xia$-dependent sub-thermal value
& bi-modally distributed between maximal and sub-thermal values
& Maximal
\\
Spectrum of local operators
& discrete
& discrete and continuous components
& continuous
\\
Local memory as $t \to \infty$ 
& yes
& on a fraction of the sites
& no

\end{tabular}
\end{ruledtabular}
\caption{\emph{Properties of avalanched systems:} Partially avalanched systems are delocalized, but non-ergodic, and do not satisfy the predictions of ETH. The table summarises the behavior of various spectral properties in the localized, partially avalanched and completely avalanched regimes. The symbol $s_{\mathrm{th }}$ denotes the thermal entropy density at the appropriate energy density from which eigenstates are drawn.}
\label{tab:nonergodic}
\end{table*}

\subsection{Partial avalanches in the Anderson Model}
\label{sec:partial_aval}

\begin{figure}
    \centering
    \includegraphics[width=0.95\columnwidth]{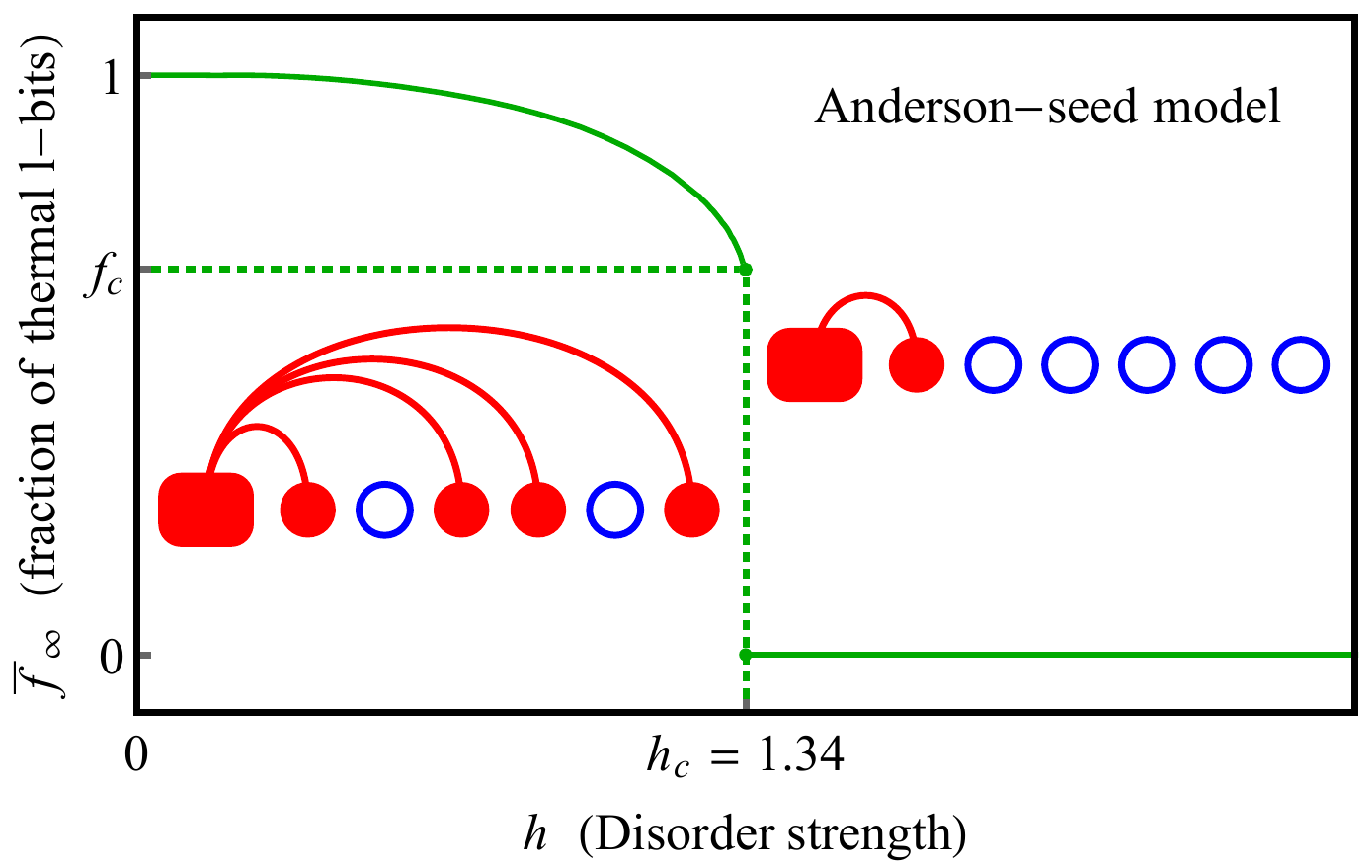}
    \caption{
    \emph{Thermal fraction of l-bits in the Anderson-seed model:} The fraction $\fA$ of l-bits absorbed into a thermal bubble as a function of disorder strength $h$ (solid green line). Insets schematically depict the resulting system. The left-most red rectangle represents the thermal seed, while the circles represent the l-bits ordered by their distance from the seed. The red l-bits participate in the thermal bubble; the red lines remind the reader that thermalization is mediated by interactions with the seed. For $h > \hc$, the thermal seed only strongly hybridises with finitely many l-bits in its vicinity. For $h<\hc$ however, a thermalization avalanche forms a non-contiguous thermal bubble of $\sim \fA L$ l-bits. 
    }
    \label{fig:PAval}

\end{figure}

When the distribution $p(\xi)$ of l-bit localization lengths has finite width (as in the Anderson chain), only a fraction of the l-bits may have long enough localization lengths by Eq.~\eqref{eq:naive} to be absorbed into the thermal bubble. The avalanche ceases when the l-bits that are not a part of the thermal bubble are too weakly coupled to the bubble to be absorbed. The stopping condition is that the inequality
\begin{equation}
    {\tilde J}_\alpha \rho_0 \sqrt{\frac{ 2^{\fA L}}{d_0 }} < \cW,
    \label{eq:frac_aval_cond}
\end{equation}
is satisfied by a fraction $(1-\fA)$ of the total number of the l-bits. 

We derive a self-consistency equation for $\fA$ below. The fraction $\fA$ is determined by the properties of the disorder distribution $\{ h_n \}$ alone and is generally between $0$ and $1$ below a critical value of the disorder strength. 

A few notational points before we proceed. Let $\xin$ ($\xim$) denote the minimum (maximum) localization length (i.e. $p(\xi) > 0$ only if $\xi \in [\xin,\xim]$). The mean localization length is given by:
\begin{align}
    [\xi] &= \int_0^\infty \xi p(\xi) d \xi = \lim_{L \to \infty} \frac{1}{L} \sum_{\alpha=1}^L \xi_\alpha \label{Eq:StatHom}
\end{align}
where $p(\xi)$ is the ensemble distribution of localization lengths.
 
Consider a small segment $\alpha/L \in [s, s + ds]$ of the disordered chain for $s \in [0,1]$ for $\alpha \gg l_{\mathrm{dir}}$. Let $\fAl(s)$ be the fraction of l-bits absorbed by the thermal bubble in this segment. This `local' fraction is related to the global fraction by integrating over $s$:
\begin{align}
 \fA = \int_0^1 \fAl(s) ds.
 \end{align}
\red{In the limit of large $L$, it follows} from Eqs.~\eqref{eq:couplings} and~\eqref{eq:frac_aval_cond} that the l-bit with index $\alpha$ does not hybridise with the bubble if
\begin{equation}
    \xi_\alpha < \frac{s \xic}{\fA}.
\end{equation}
The local fraction of l-bits which satisfy the above condition is thus:
\begin{equation}
    1-\fAl(s) = \int_{0}^{s \xic/\fA} d\xi' \, p(\xi').
    \label{eq:fAs}
\end{equation}
Integrating over $s$, we obtain an equation which determines $\fA$ implicitly
\begin{align}
\fA & = 1 - \int_{0}^1 ds \int_{0}^{s \xic/\fA} d\xi' \, p(\xi')  \nonumber
\\
& = 1 - \int_0^{\xic / \fA} d \xi' \, p(\xi') \left( 1- \frac{ \fA \xi'}{\xic} \right).
\label{Eq:SelfconsistentfA}
\end{align}
Solutions to Eq.~\eqref{Eq:SelfconsistentfA} determine the fraction of l-bits absorbed into the thermal bubble. These solutions define three distinct regimes. We describe these regimes below, and delegate some mathematical details to Appendix~\ref{app:partial_proofs}.

\begin{itemize}
    \item \emph{No avalanches for sufficiently localized chains:} A trivial solution to Eq.~\eqref{Eq:SelfconsistentfA} is $\fA=0$. The thermal bubble then fails to absorb a non-zero fraction of the l-bits in the chain. The stability of the $\fA=0$ solution determines whether or not the l-bits are asymptotically localized for arbitrarily large thermal seeds. For $\xia < \xic$, $\fA=0$ is the only solution and is stable:
\begin{equation}
    \fA = 0 \quad \textrm{for} \quad  \xic > \xia := \int_0^\infty d \xi' \, \xi' \, p(\xi')
\end{equation}
    Thus, when $\xia<\xic$, there is no avalanche instability. The condition $\xia = \xic$ determines the avalanche threshold. Below threshold $h < \hc$ ($\xia > \xic$), $\fA = 0$ remains a solution, but is unstable.
    
    \item \emph{Complete avalanches for sufficiently delocalized chains:} When the localization lengths $\xi_\alpha$ are all larger than the threshold value $\xic$, all l-bits are absorbed into the thermal bubble
\begin{equation}
    \fA = 1 \quad \textrm{for} \quad \xin > \xic.
\end{equation}
    This is the DRH result of Sec.~\ref{Sec:DRHTheory}. 
    
    \item \emph{Partial avalanches for generic distributions}: When $\xin<\xic<\xia$, there is a stable solution to Eq.~\eqref{Eq:SelfconsistentfA} with a non-zero $\fA$ satisfying $\fA \geq \fAc := \xic/\xim$:
\begin{equation}
    \fAc \leq \fA < 1 \quad \textrm{for} \quad \xin < \xic < \xia.
    \label{Eq:fAgeneralform}
\end{equation}
    As $\fA<1$, the resulting avalanche is only partial. At late times, a non-contiguous \emph{thermal bubble} comprising a fraction $\fA$ of the l-bits co-exists with the remaining localized l-bits. \redd{This intermediate partially avalanched regime is generic: if $p(\xi)$ has any finite width as $\xia$ is tuned through the critical value $\xic$, then Eq.~\eqref{Eq:fAgeneralform} will be satisfied.}
\end{itemize}

Fig.~\ref{fig:PAval} shows the dependence of $\fA$ on the disorder strength $h$ for a model with the same $p(\xi)$ as the Anderson model in Eq.~\eqref{Eq:AndersonLbitHam}. 
The $p(\xi)$ for the Anderson model with box disorder was obtained numerically, see Appendix~\ref{app:anderson_analysis}.
Numerically, the disorder strength and thermal bubble fraction at threshold take the values:
\begin{equation}
\begin{aligned}
    \hc \approx 1.37 \quad \text{and} \quad \fAc \approx 0.67.
    \label{eq:anderson_hc_fc}
\end{aligned}
\end{equation}
The inset of Fig.~\ref{fig:PAval} schematically depicts the spatial structure of the thermal bubble below threshold.

A number of comments are in order. First, Fig.~\ref{fig:PAval} shows the unique stable solution to Eq.~\eqref{Eq:SelfconsistentfA}. Although $\fA=0$ is a solution at any $h$, its instability below threshold indicates that if an avalanche starts, it forms a thermal bubble with $\sim \fA L \geq \fAc L $ l-bits. Second, the $\fA$ vs $h$ curve is discontinuous at threshold if and only if the disorder distribution has bounded support. That is, $\fAc$ is non-zero if and only if $\xim < \infty$, as is the case for the Anderson model with box disorder. Third, the fraction of l-bits absorbed into the thermal bubble within a segment $\alpha/L \in [s,s+ds]$ (given by $f_\infty(s)$) can exhibit strong dependence on the distance from the seed. For example, Eq.~\eqref{eq:fAs} implies that $\fAl(s=0) = 1$ and $\fAl(s=1) \leq 1$ for $h<\hc$. Most dramatically, $\fAl(s=1) \to 0$ as $h \to \hc^-$, i.e. the l-bits at the end of the chain are always localized at threshold.

The partially avalanched phase is thus an example of a delocalized non-ergodic phase. A number of spectral and dynamical properties immediately follow from the simple picture of a thermal bubble with $\sim \fA L$ l-bits co-existing with localized l-bits. These are summarised in Table~\ref{tab:nonergodic}. We return to these spectral properties in Sec.~\ref{sec:numerics}, when we numerically test the predictions of the extended avalanche theory for an Anderson chain coupled to a thermal seed.

\subsection{Avalanches in the Random-Heisenberg model}
\label{Sec:AvalanchesRandomHeisenberg}
The key difference between an MBL system and an Anderson chain coupled to a single thermal seed is that, in the MBL chain, for \redd{$\xia > \xic$}, the avalanche instability leads to complete thermalization irrespective of the distribution of l-bit localization lengths.
This is because an MBL system has a finite density of rare low-disorder regions of any given size, and thus a finite density of thermal seeds.
As the coupling strength between any l-bit and the closest seed participating in the bubble $\tilde{J}_\alpha$ is finite as $L \to \infty$, the LHS of the inequality:
\begin{align}
{\tilde J}_\alpha \rho_0 \sqrt{\frac{ 2^{\fA L}}{d_0 }} \gg \cS
\end{align}
is exponentially larger than the RHS and every l-bit is absorbed into the thermal bubble. 

Nevertheless, it may be that the distribution of localization lengths plays a role in analytical predictions about the MBL transition based on avalanches~\cite{thiery2018many}.
The numerical study of Ref.~\cite{varma2019length} suggests that deep in the MBL phase, the l-bits of the random-Heisenberg model are characterised by a unique localization length.
Ascertaining whether the localization length is necessarily unique at intermediate strengths near the putative MBL transition is an interesting topic for future study.

\begin{figure*}
    \centering
    \includegraphics[width=0.8\textwidth]{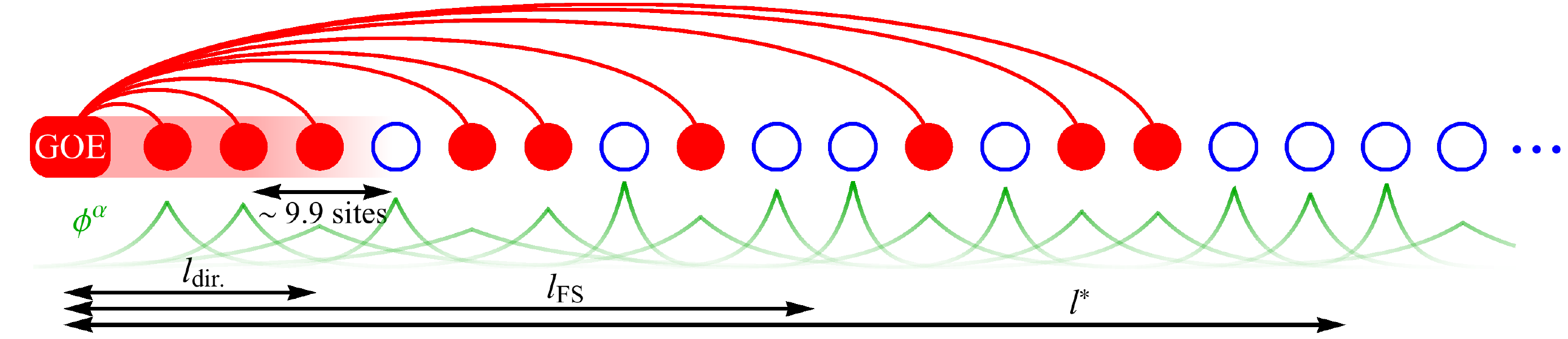}
    \caption{\emph{Schematic avalanche:} The thermal seed (red rectangle) and the red l-bits form a non-contiguous thermal bubble. The blue l-bits have short localization length and are weakly coupled to the thermal bubble. Direct coupling to the seed adds $l_{\mathrm{dir}}$ l-bits to the thermal bubble (red shaded area). The boundary to the direct coupling region is spread over $\approx 9.9$ sites. Below the Harris-Luck scale $l_{\mathrm{FS}}$, fluctuations in the window-averaged disorder make it impossible to determine whether the system is above or below the avalanche threshold. The thermal bubble has to grow to a size comparable to $l^\star$ to cause runaway thermalization. In the figure, $l_{\mathrm{dir}} < \textrm{max}(l^\star, l_{\mathrm{FS}})$ and the thermal bubble is typically of finite extent. 
    }
    \label{fig:Lengths}

\end{figure*}

\section{Avalanches at finite $l_{\mathrm{dir}}$}
\label{sec:threshold}

The number of l-bits that are directly thermalized by the bare thermal seed, $l_{\mathrm{dir}}$, controls the probability of an avalanche starting below threshold.
If $l_{\mathrm{dir}}$ is too small, then only rare sample will avalanche and $\fbig \approx 0$ irrespective of the value of $\xia$ (see Fig.~\ref{Fig:CriticalFan}).

In this section, we derive the minimum value of $l_{\mathrm{dir}}$ such that typical samples avalanche below threshold and $\fbig(h,l_{\mathrm{dir}}) \approx \fA(h)$ for $h<\hc$.
We find that the minimum value is set by the maximum of two length scales, $l^\star$ and $l_{\mathrm{FS}}$, so that:
\begin{align*}
l_\mathrm{dir} &> \max( l^\star , l_{\mathrm{FS}}) \quad  \textrm{typical samples avalanche} \\
l_\mathrm{dir} &< \max( l^\star , l_{\mathrm{FS}}) \quad \textrm{typically finite thermal bubble}
\end{align*}
We show that $l^\star$ and $l_{\mathrm{FS}}$ diverge with different exponents as $h \to \hc$.
Consequently, the probability to start an avalanche with fixed $l_\mathrm{dir}$ rapidly (exponentially) goes to zero as threshold is approached $h \to \hc^-$, and $\fbig(h, l_\mathrm{dir})$ rapidly approaches zero as $h \to \hc^-$ (see Fig.~\ref{fig:Scales}(b)). In comparison, $\fA(h)$ jumps at $h=\hc$ (see Fig.~\ref{fig:PAval}).
The phase diagram is summarised in Fig.~\ref{Fig:CriticalFan}.

We note that multiple diverging length scales near a dynamical transition in a disordered system is reminiscent of infinite-randomness physics.
We return to this point in the discussion section.

\subsubsection{$l_\mathrm{dir}$: locus of direct thermalization due to the seed}
\label{sec:ldir}
L-bits with index $\alpha \leq l_\mathrm{dir}$ are sufficiently strongly coupled to the seed so as to directly hybridise with it. Thus, $l_\mathrm{dir}$ sets the scale to which the thermal bubble is guaranteed to grow. As the weak and strong coupling thresholds, $\cW$ and $\cS$, are separated by over an order of magnitude, it is not possible to define this length precisely. However, by the arguments in Sec.~\ref{sec:runaway_theory}, the length is bounded by the relation
\begin{equation}
    \cW \leq \frac{\tilde{J}_{l_\mathrm{dir}} \rho_0}{\sqrt{d_0}} \leq \cS.
\end{equation}
Using $|\phi_1^\alpha| \sim \e^{-\alpha / \xi_\alpha}$ and considering a typical sample (ie taking $\xi_\alpha \approx [\xi]$ in the vicinity of the seed) we obtain: 
\begin{equation}
    l_\mathrm{S} \leq l_\mathrm{dir} \leq  l_\mathrm{W} = l_\mathrm{S} + \xia \log \frac{\cS}{\cW}.
    \label{eq:ldir}
\end{equation}
where the length scale $l_\mathrm{S}$ sets the range of l-bits strongly coupled to the seed
\begin{equation}
    l_\mathrm{S/W} : = \xia\log \left( \frac{J \rho_0}{2 c_{\mathrm{S/W}} \sqrt{d_0} } \right).
    \label{eq:LSW}
\end{equation}
As the avalanche threshold is approached $\xia \to \xic$ the length scale $l_\mathrm{dir}$ remains finite and specified by $l_\mathrm{S} \leq l_\mathrm{dir} \leq  l_\mathrm{S} + 9.9$.

\subsubsection{$l^\star$: a length scale controlling runaway thermalization}

Define $p_\mathrm{stop}(l)$ to be the probability that a thermal bubble with $\sim \fA l$ l-bits is stable to the inclusion of more l-bits, i.e. that the thermal bubble `stops' at length $l$. For $h<\hc$, this probability is exponentially decaying. We define this decay length to be $l^\star$
\begin{equation}
    p_\mathrm{stop}(l) \sim \e^{- l / l^\star}.
    \label{eq:pstop}
\end{equation}
A thermal bubble of length $l > l^\star$ has a finite probability of absorbing $O(L)$ l-bits. Below, we derive an expression for $l^\star$, and discuss the implications of its divergence near threshold. 

Consider a partially avalanched Anderson chain below threshold ($h < \hc$) of length $l \gg 1$. By Eq.~\eqref{Eq:SelfconsistentfA}, the thermal bubble has $\sim \fA l$ l-bits. Suppose we increase the length of chain to some $L \gg l$. If the $(L-l)$ l-bits in the extended part of the chain are too weakly coupled to the bubble, then the thermal bubble's size remains $\sim \fA l$. The condition for weak coupling is:
\begin{equation}
    \tilde{J}_\alpha \rho_0 \sqrt{\frac{2^{\fA l}}{d_0}} < \cW \quad \forall \alpha > l.
\end{equation}
By taking $\tilde{J}_\alpha = \tfrac12 J \phi_1^\alpha \approx \tfrac12 J \e^{-\alpha/\xi_\alpha}$ and rearranging, we find that the thermal bubble's size is self-consistently $\sim \fA l$ if:
\begin{equation}
    \xi_\alpha < \frac{\alpha }{ \fA l / \xic + l_{\mathrm{W}}/\xia} \quad \forall \alpha > l.
    \label{eq:stopping}
\end{equation}
Here $l_{\mathrm{W}}$ (Eq.~\eqref{eq:LSW}) sets the number of l-bits with strong or intermediate coupling to the seed.

The probability that Eq.~\eqref{eq:stopping} is satisfied for all $\alpha>l$ is given by the product of the probabilities that each l-bit satisfies Eq.~\eqref{eq:stopping}~\footnote{ We note that Eq.~\eqref{eq:pstop_prod} is exact only if nearby $\xi_\alpha$ are independent and identically distributed. In the Anderson chain, nearby $\xi_\alpha$ are correlated. Nevertheless, as the correlations decay exponentially in the distance between the l-bits, the correction is $\mathrm{O}(l^0)$ in the exponent of Eq.~\eqref{eq:pstop}.}. Thus, the stopping probability for a thermal bubble with $\sim \fA l$ l-bits is:
\begin{equation}
\begin{aligned}
    p_\mathrm{stop}(l) &= \prod_{\alpha = l + 1}^{L} \int_0^{\alpha/(\fA l/\xic + l_{\mathrm{W}}/\xia)} d \xi' p(\xi').
\end{aligned}
\label{eq:pstop_prod}
\end{equation}
The number of terms in the product which are less than unity (i.e. those with $\alpha < \xim( \fA l / \xic+ l_{\mathrm{W}}/ \xia )$) scales with $l$, and hence the probability that the thermal bubble stops growing at size $l$ is exponentially small for $l > l^\star$ (Eq.~\eqref{eq:pstop}). 

\begin{figure}
    \centering
    \includegraphics[width=0.95\columnwidth]{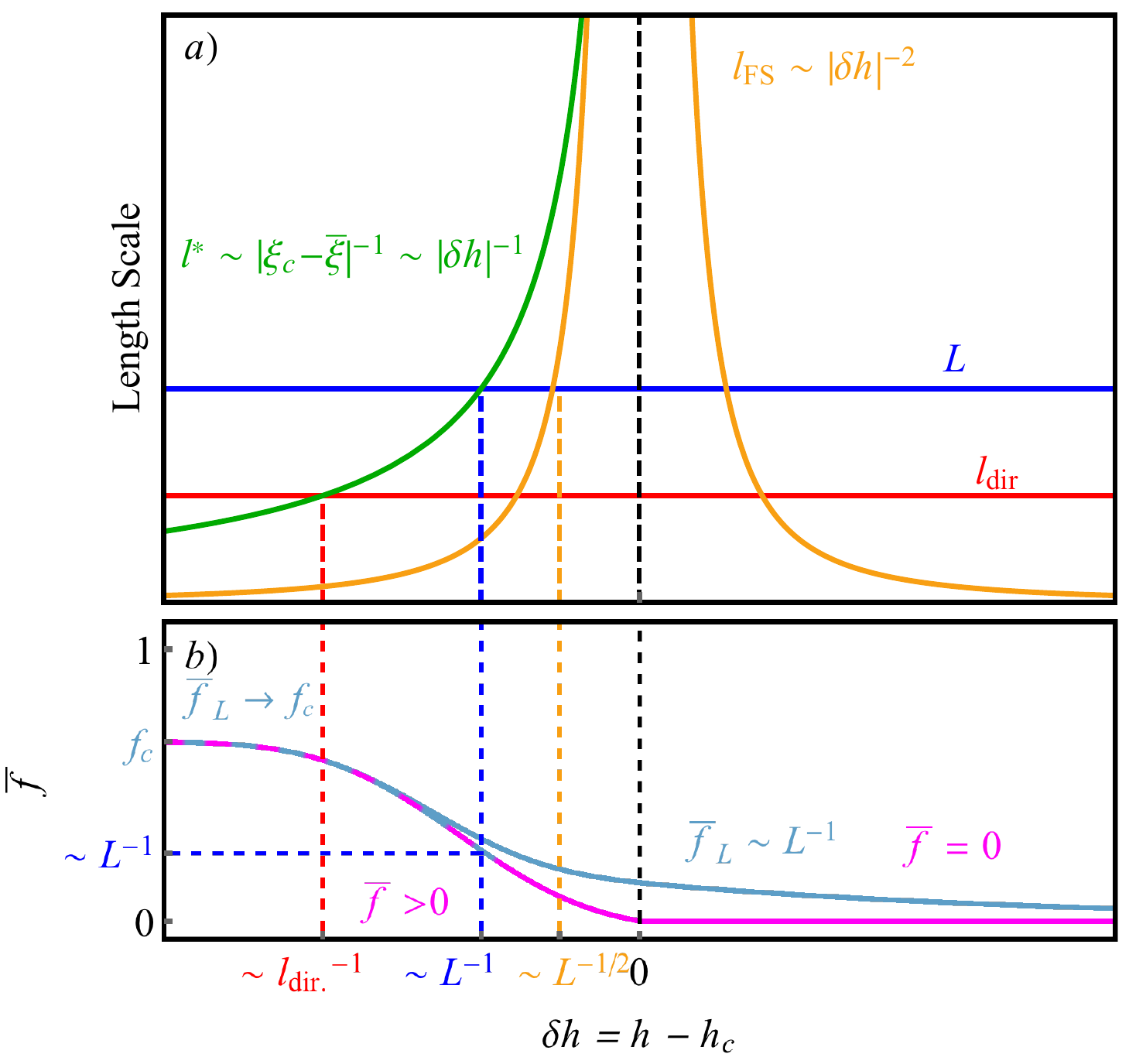}
    \caption{\emph{Length scales near the avalanche threshold:} The upper plot shows the instability scale $l^\star$ (green), and the Harris-Luck scale $l_\mathrm{FS}$ (yellow) diverging as $\delta h \to 0$, while the length of the chain $L$ (blue) and the locus of direct thermalization $l_\mathrm{dir}$ (red) remain finite. The lower plot depicts $\fbig$ as a function of $\delta h$ for fixed $l_\mathrm{dir}$ in the thermodynamic (magenta) and finite chain (cyan).  The mean thermodynamic fraction $\fbig$ deviates from $\fAc$ when $l^\star$ becomes comparable to $l_\mathrm{dir}$ and approaches zero as $\delta h \to 0^-$. At finite values of $L$ (cyan), $\fbig_L$ saturates to an $O(L^{-1})$ value when $l^\star > L$ or $\delta h \sim L^{-1}$. This $O(L^{-1})$ value decreases with increasing $\delta h$. At larger $L$ however, the blue dashed line is to the right of the yellow dashed line, and $\fbig_L$ saturates to an $O(L^{-1})$ value when $l_\mathrm{FS} > L$.
    }
    \label{fig:Scales}
\end{figure}

In Appendix~\ref{app:pstop}, we show that $l^\star$ diverges as the avalanche threshold is approached from the avalanching phase: 
\begin{equation}
    l^\star \sim \frac{\xic}{\fAc(\xia - \xic)} \propto |h - \hc|^{-1}
    \label{eq:critical_lstar}
\end{equation}
Above $\sim$ indicates asymptotic equality. The proportionality to $| h - \hc|^{-1}$ follows as $\xia$ is a continuous function of $h$ with finite gradient at the avalanche threshold. The diverging length scale $l^\star$ is shown in green in Fig~\ref{fig:Scales}a.

For $h<\hc$, the chain typically avalanches if $l^\star < l_\mathrm{dir}$. The resulting thermal bubble has $\sim \fA L$ l-bits and is described by the partial avalanche theory of Sec.~\ref{sec:partial_aval}. This regime is depicted to the left of the dashed red line in Fig.~\ref{fig:Scales}b.

For $h<\hc$ and $l^\star > l_\mathrm{dir}$, the thermal seed typically stops after absorbing $\mathrm{O}(l_\mathrm{dir})$ l-bits. This regime occurs between the red and blue dashed lines in Fig.~\ref{fig:Scales}b. As runaway thermalization occurs in atypical samples where the thermal seed absorbs of order $l^\star$ l-bits, the mean fraction of l-bits in the thermal bubble satisfies the following inequality:
\begin{align}
0 \leq \fbig(h, l_\mathrm{dir}) < \fA(h), \quad l^\star \gg l_\mathrm{dir}
\end{align}
As $h \to \hc^-$, stopping probability in Eq.~\eqref{eq:pstop_prod} becomes non-decaying with $l$ and consequently $\fbig$ approaches zero (magenta line in Fig.~\ref{fig:Scales}b).

Let us turn to the effects of a finite chain length $L$. 
At finite $L$, there is a regime near threshold where $ l^\star > L > l_\mathrm{dir}$. 
In this regime, the stopping probability $p_\mathrm{stop}$ is effectively constant over the length of the chain. The thermal bubble then typically grows to a size of order $1/|\log p_\mathrm{stop}|$ before stopping. Consequently, the fraction of the l-bits in the chain that typically participate in the thermal bubble decreases with $L$. Using the relationship between $l^\star$ and disorder strength near threshold (Eq.~\eqref{eq:critical_lstar}):
\begin{equation}
   \fbig(\hc , l_\mathrm{dir}) = \mathrm{O}(L^{-1}), \quad |h - \hc| \lesssim L^{-1}
\end{equation}
See the cyan curve in Fig~\ref{fig:Scales}b.

\redd{
In this argument we have treated the localisation lengths of orbitals as iid drawn from the marginal distribution $p(\xi)$. In reality there are correlations: nearby levels repel in energy, and as the localisation length is a function of the orbital energy only, there are corresponding correlations in the localisation lengths of nearby levels. However these correlations decay exponentially on a characteristic length set by the localisation lengths of the orbitals in question. As the localisation length is asymptotically smaller than $l^\star$ as the avalanche threshold is approached, the short range correlations are irrelevant to the argument presented here.
}

Thus far, we have ignored fluctuations of the disorder strength on the scale of the thermal bubble, i.e. we have implicitly assumed that $l, l^\star \gg l_\mathrm{FS}$. This assumption is manifest in the appearance of the global mean localization length $\xia$ in the definition of $l^\star$. Sufficiently close to threshold, the windowed average of the disorder shows significant spatial fluctuations, and the assumption that $l, l^\star \gg l_\mathrm{FS}$ is violated. We derive $l_\mathrm{FS}$ next.

\subsubsection{$l_\mathrm{FS}$: the Harris-Luck scale}
\label{sec:LFS}

The Harris-Luck scale~\cite{harris1974effect,chayes1986finite,luck1993classification,chandran2015finite} $l_\mathrm{FS}$ sets the minimum size of the window required to accurately estimate the average disorder strength in the chain.
As the average disorder strength controls runaway thermalization (see Fig.~\ref{fig:PAval}), quantities such as $\fbig$ which can differentiate between the avalanched and localized phases cannot be accurately determined on length scales much lesser than $l_\mathrm{FS}$.

A precise definition of $l_\mathrm{FS}$ is as follows. Consider the localization length averaged over the l-bits within a spatial window of length $l$:
\begin{equation}
    \bar{\xi}_l = \frac{1}{l} \sum_{\alpha = 1}^{l} \xi_\alpha
\end{equation}
The windowed average $\bar{\xi}_l$ varies from one sample to the other (or equivalently, with the position of the window within an infinite sample). 
The deviation from the ensemble mean is set by the central limit theorem:
\begin{equation}
    |\bar{\xi}_l - \xia | \sim \sqrt{[ (\bar{\xi}_l - \xia)^2 ] }  \sim \frac{\sigma(\xi)}{\sqrt{l}},
\end{equation}
where $\sigma(\xi) = [\xi^2]-[\xi]^2$ and $[\bar{\xi}_l] = \xia$. 

If the deviation of the windowed average from the ensemble mean is greater than the difference between $\xia$ and $\xic$, no window averaged `order parameter' can consistently determine whether the chain is above or below threshold.  $l_\mathrm{FS}$ is defined as the threshold window size at which the deviation is comparable to $|\xia - \xic |$:
\begin{align}
    |\bar{\xi}_{l_\mathrm{FS}} - \xia | &= |\xia - \xic| \\
   \Rightarrow l_\mathrm{FS} &= \frac{\sigma^2(\xi)}{|\xia - \xic|^{2}} \sim  | h - \hc|^{-2}.
\end{align}

Close to the avalanche threshold, $l_\mathrm{FS}$ sets minimum size of a thermal bubble that can cause runaway thermalization. This follows from the corrected version of Eq.~\eqref{eq:pstop_prod} with $\xia$ replaced by $\bar{\xi}_{al}$ for $l < l_\mathrm{FS}$. Asymptotically, we then find:
\begin{align}
    p_\mathrm{stop}(l) \sim \exp\left( -  l \frac{\fAc(\bar{\xi}_{(a l)}-\xic)}{\xic} \right),
    \label{eq:pstop_flucs}
\end{align}
where $a$ is an $O(1)$ number. By the central limit theorem 
\begin{equation}
    \bar{\xi}_{(a l)} = \xia + \eta_l \sigma(\xi)/\sqrt{a l}    
    \label{eq:eta}
\end{equation}
where $\eta_l$ is a standard normal random variable $[\eta_l] = 0$, $[\eta_l^2] = 1$. Substituting~\eqref{eq:eta} into~\eqref{eq:pstop_flucs} we obtain
\begin{align}
    p_\mathrm{stop}(l) = \exp\left( - \frac{l}{l^\star} - \eta_l \frac{\fAc \sigma(\xi)}{\xic} \sqrt{\frac{l}{a}} \right).
\end{align}
When the second term in the exponent is comparable to the first term, the two terms may cancel and produce significant stopping probability. Such cessation of the avalanche due to such local fluctuations in the disorder cannot happen once the first term dominates
\begin{equation}
    l \gg l_\mathrm{FS} = \frac{\sigma^2(\xi)}{a(\xia - \xic)^2} \sim |h-\hc|^{-2}
\end{equation}

Close to threshold, when $l^\star \ll l \lesssim l_\mathrm{FS}$, the growing thermal bubble will absorb typical regions. However, it will also encounter rare regions where the Anderson chain appears locally to be more localized. Such rare regions can stop the growth of the bubble, and hence stop the avalanche. This mode of failure due to rare regions becomes exponentially unlikely once the thermal bubble grows to a length scale $l > l_\mathrm{FS}$. As $l_\mathrm{FS}$ diverges near threshold, the probability of an avalanche starting goes to zero (magenta curve in Fig~\ref{fig:Scales}), and the thermal bubble size is typically $l_{\mathrm{dir}}$.

Finally, we emphasize that $l_\mathrm{FS}$ controls finite size scaling near the avalanche threshold for large enough $L$~\cite{harris1974effect,chayes1986finite,luck1993classification,chandran2015finite}. Fig.~\ref{fig:Scales} depicts a pre-asymptotic regime in which the finite-size corrections are set by $l^\star$. We depict this pre-asymptotic regime as it  is likely that numerically accessible chains are in this regime.

\section{Observing the avalanche instability}
\label{sec:runaway_conditions}
The requisite hierarchy of length scales to conclusively observe the avalanche instability in typical samples below the avalanche threshold is $  l^\star, l_\mathrm{FS} \ll l_\mathrm{dir} \ll L$. In this section, we identify parameter regimes of the Anderson-seed chain in Fig.~\ref{fig:Cartoon} in which this hierarchy holds to test our extended avalanche theory. Numerically however, the chain lengths are so short that $l_{\mathrm{dir}}$ is comparable to $L$.

These conditions supplement those of Sec.~\ref{sec:starting} which derived conditions on the thermal seed so that the first $l_\mathrm{S}$ l-bits strongly hybridises with the seed. That is,  Sec.~\ref{sec:starting} asked that $l_\mathrm{dir}> l_\mathrm{S} >0$. 

\subsection{$L > l_\mathrm{dir}$ condition: Furthest l-bits are not thermalized by direct processes}
\label{sec:length_condition}

To numerically determine whether l-bits are thermalized by avalanches or by direct coupling with the seed requires sufficiently long Anderson chains. That is, $L$ should exceed $l_\mathrm{dir}$. Using the inequality derived in Sec.~\ref{sec:ldir}:
\begin{equation}
    l_\mathrm{S} < l_\mathrm{dir} < l_\mathrm{S} + \xia \log \frac{\cS}{\cW}.
\end{equation}
we find:
\begin{equation}
    L > l_\mathrm{S} + \xia \log \frac{\cS}{\cW} > 1 + \xic \log \frac{\cS}{\cW} = 10.9\ldots.
    \label{eq:size_cond}
\end{equation}
Above, we assumed that the chain is in the avalanching regime ($\xia > \xic$), and that $l_\mathrm{S} \geq 1$ (see Sec.~\ref{sec:goodthermalseed}).

\subsection{Partial avalanche condition: dominant fraction of l-bits not at intermediate couplings}
\label{sec:partial_conditions}

Localized l-bits co-exist with a macroscopically large thermal bubble in the partially avalanched regime. The co-existence results in a bi-modal orbital/single-site entanglement entropy distribution obtained by tracing out each l-bit/site in eigenstates. The bi-modality quantifies the non-ergodic nature of the partially avalanched phase (see Table~\ref{tab:nonergodic}).

At finite $L$, the bi-modal signature can be washed out by the l-bits that have intermediate coupling to the \emph{bubble} and hence intermediate orbital entropy values. The condition for intermediate coupling is:
\begin{equation}
    \frac{\cW}{c_0 \sqrt{2^{\fA L}}} < \phi_1^\alpha < \frac{\cS}{c_0 \sqrt{2^{\fA L}}},
    \label{eq:intermediate_coupling}
\end{equation} 
where $c_0 = \tfrac12 J \rho_0 / \sqrt{d_0}$. This condition follows from the discussion around Eq.~\eqref{eq:frac_aval_cond}. 

As $\phi_1^\alpha$ decays exponentially with $\alpha$, the number of intermediately coupled l-bits is $\mathrm{O}(1)$, as compared to the $\mathrm{O}(L)$ number of localized l-bits.
Nevertheless, at finite $L$, the former can exceed the latter.
Below, we numerically compute the $L$ at which the two are equal in the Anderson chain.

\begin{figure}[!]
    \centering
    \includegraphics[width=0.95\columnwidth]{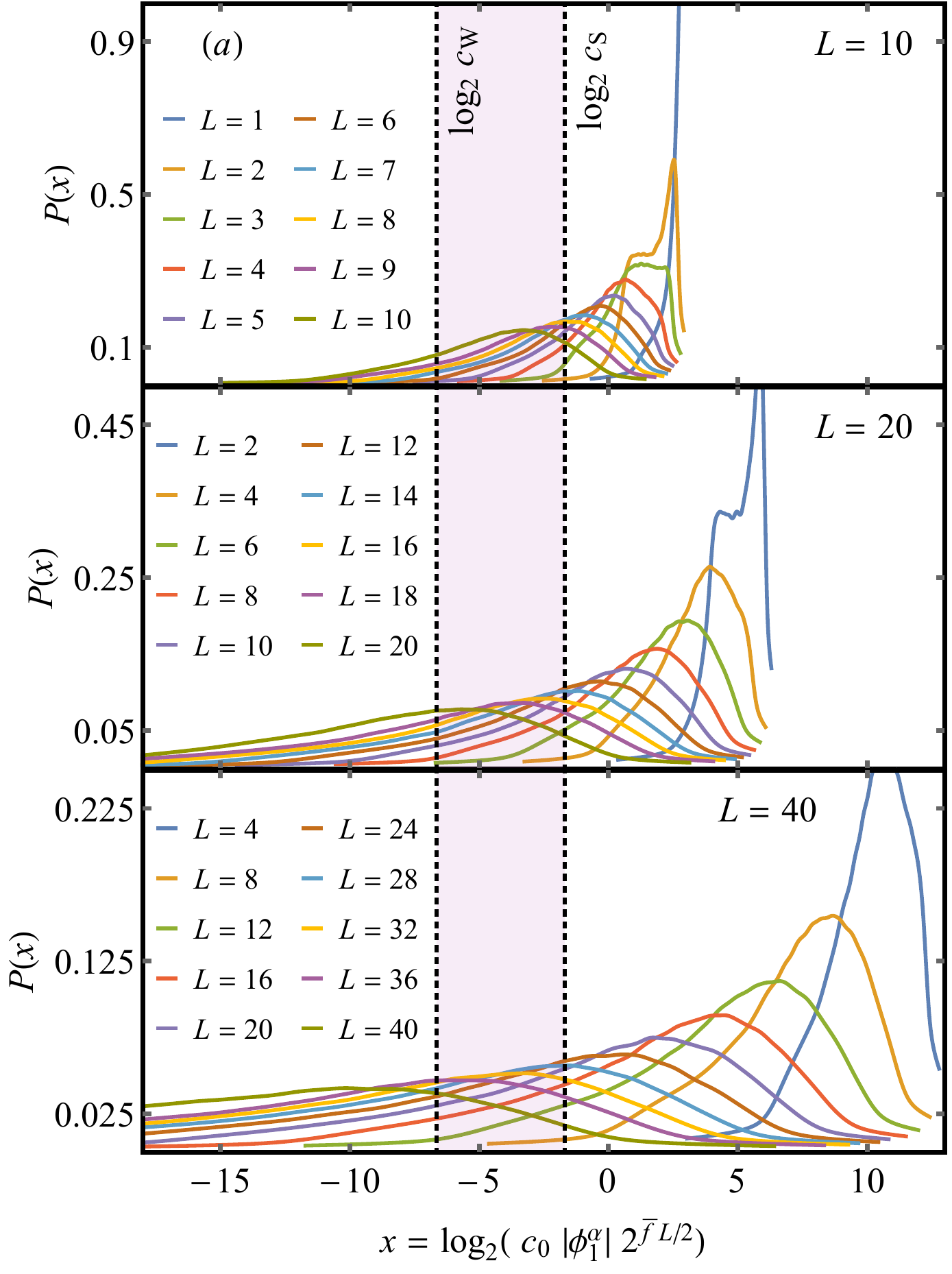}
    \includegraphics[width=0.95\columnwidth]{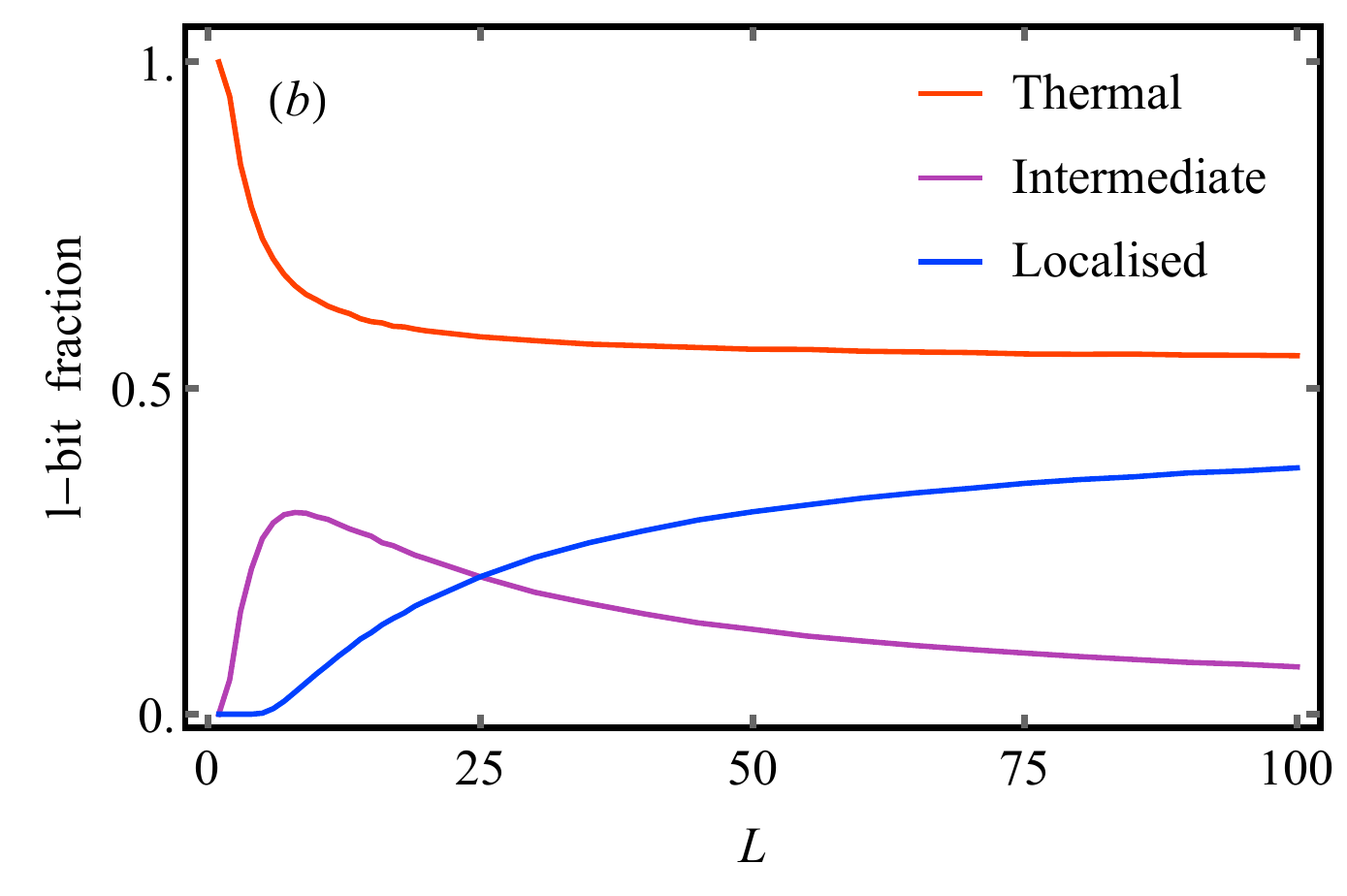}
    
    \caption{
    \emph{Finite size effect due to intermediate couplings between the l-bits and the thermal bubble:} Panel (a) shows the distributions of the quantity $x = \log_2 (c_0 |\phi_1^\alpha| \sqrt{2^{\fbig L}})$ for $|\phi_1^\alpha|$ obtained from exact diagonalization of an open Anderson chain. Data is shown for different $\alpha$ (legends inset) and different chain lengths. L-bits with intermediate coupling to the thermal bubble fall in the violet region. (b) The total fraction of l-bits lying in the intermediate region (violet line) scales as $L^{-1}$. This fraction is lesser than the total localized fraction (blue line) at $L \approx 26$. Parameters: $h=\hc$, $\fbig = \fAc$ from Eq.~\eqref{eq:anderson_hc_fc}.
    }
    \label{fig:LLflucs}

\end{figure}

Fig~\ref{fig:LLflucs}a shows the disorder averaged distributions of $x = \log_2 (c_0 |\phi_1^\alpha| \sqrt{2^{\fA L}})$ for $h=\hc$ at different $L$. The orbital index is shown in the legend.
The violet region marks the intermediate coupling regime in Eq.~\eqref{eq:intermediate_coupling}. 

Orbitals for which $x$ falls to the right of the violet region in the thermal region ($x > \log \cS$) are strongly coupled to the thermal bubble and avalanche theory predicts that they will be absorbed into the thermal bubble. Those to the left in the localized region ($x < \log \cW$) are only perturbatively corrected by the thermal bubble. As the distributions of $x$ widen and decrease in height with increasing $L$, the fraction of l-bits in the intermediate regime decrease with increasing $L$.

The disorder averaged total fraction of l-bits in the thermal, intermediate and localized regions are plotted in Fig~\ref{fig:LLflucs}b. 
As expected, the fraction of intermediately coupled l-bits falls as $\mathrm{O}(L^{-1})$. However, the fraction of intermediate l-bits falls below the fraction of localized l-bits only at $L \approx 26$. Thus, to see clear signatures of the predicted bi-modality of single-site eigenstate entanglement entropy requires:
\begin{equation}
    L \gtrsim 26.
    \label{eq:L26}
\end{equation}

\subsection{Other conditions}
As $l_\mathrm{dir}$ could be as large as $l_\mathrm{S}+11$ sites, the numerics described below with $L\approx 10$ is likely in the regime $l_\mathrm{dir} \lesssim L$. We therefore do not identify further parameter regimes based on the hierarchy between $l_\mathrm{dir}$ and $l^\star$ or $l_{\mathrm{FS}}$.

\section{Numerical evidence for partial avalanches}
\label{sec:numerics}
We present numerical tests which support the partial avalanche theory in the Anderson-seed model of Eq.~\eqref{eq:fullH}. Specifically, we find that below a threshold disorder strength (i)  l-bits with exponentially small in $L$ bare couplings to the seed have thermal eigenstate expectation values, (ii) the matrix-element-to-level-spacing ratio of an operator on the thermal seed is exponentially large in $L$, and \red{(iii) the nearest neighbour level spacing ratio $\bar{r}$ flows towards the Poisson value $\bar{r} = 2 \log 2 - 1$ with increasing $L$ in all cases. These results are explained by the presence of a thermal bubble with $\sim \fbig L < L$ l-bits.}

\red{
We introduce a new statistical indicator (denoted by $v$) in Sec.~\ref{sec:v} which directly probes the effective size of the thermal bubble, and thus detects the non-ergodic nature of the delocalization in the avalanched regime. It is also insensitive to the spatial inhomogeneity of the thermal bubble. In contrast, standard spectral signatures such as the half-cut entanglement entropy (at late times or in eigenstates) and the energy level spacing that detect violations of the ETH are difficult to interpret conclusively. The behavior of standard spectral signatures is discussed in Table~\ref{tab:nonergodic}.
}

\red{Finally, we present numerical tests of the Heisenberg-seed model in Sec.~\ref{Sec:NumericsHeisenberg}. As discussed in Sec.~\ref{Sec:AvalanchesRandomHeisenberg}, interacting fermionic chains have a finite density of low-disorder regions, and thus a finite density of thermal seeds. We therefore expect that (i) MBL is stable if and only if $\xia<\xic$, (ii) the avalanche instability when $\xia>\xic$ leads to complete thermalization and an ETH phase, and (iii) the $L \to \infty$ dynamical phase diagram is unaffected by the inclusion of a single finite thermal seed. Nevertheless, our numerical study in Sec.~\ref{Sec:NumericsHeisenberg} has three aims. The first is to demonstrate that the $v$-statistic introduced in Sec.~\ref{sec:v} detects the MBL-thermal transition as a function of disorder strength. The second is show that thermalization is complete in the delocalized phase by measuring the effective size of the thermal bubble to be $\sim L$. The third is to test whether the critical disorder strength shifts to larger values due to the inclusion of a finite thermal seed in the Heisenberg model at numerically accessible system sizes. We however find no evidence of such a shift within the accuracy of our numerics.}

\subsection{Parameter regime}

We set $J=1$, $\Delta = 0$, $d_0 = 16$, and $\WB = 0.8$, and present data on either side of the avalanche threshold $\hc \approx 1.37$. These parameters satisfy the conditions laid out in sections~\ref{sec:starting} and~\ref{sec:runaway_conditions}.

Precisely, we satisfy the Bandwidth condition Eq.~\eqref{eq:condition_bandwidth}
\begin{equation}
    4 \WB = 3.2 > J + h = 2.37;
\end{equation}
and the coupling strength condition Eq.~\eqref{eq:condition_coupling} 
\begin{equation}
    \frac{\tilde{J}_1 \rho_0}{ \sqrt{d}} \approx 0.45 > \cS = 0.31.
\end{equation}
Above, we have used $J_1 = \tfrac12 J \psi_1^1$, $\psi_1^1 \approx 1/\sqrt{\xi}$, $\xi \approx \xic$, and the GOE density of states at maximum entropy $\rho_0 = d_0/(\pi \WB)$. We have numerically checked that the first l-bit is strongly hybridised with the thermal seed at these parameter values (data not shown).

We note the chain length condition, Eq.~\eqref{eq:size_cond}, is $L > 10.9$ for the observation of avalanches. Furthermore, we expect significant bi-modality in the distribution of eigenstate expectation values only for $L \gtrsim 26$, (see Section~\ref{sec:partial_conditions}). Consequently, for numerically accessible $L$, we indeed see no evidence for bi-modality.

\subsection{Thermalization of l-bits with exponentially weak coupling to the thermal seed}
\label{Sec:Sphistar}

\begin{figure}
    \centering
    \includegraphics[width=0.95\columnwidth]{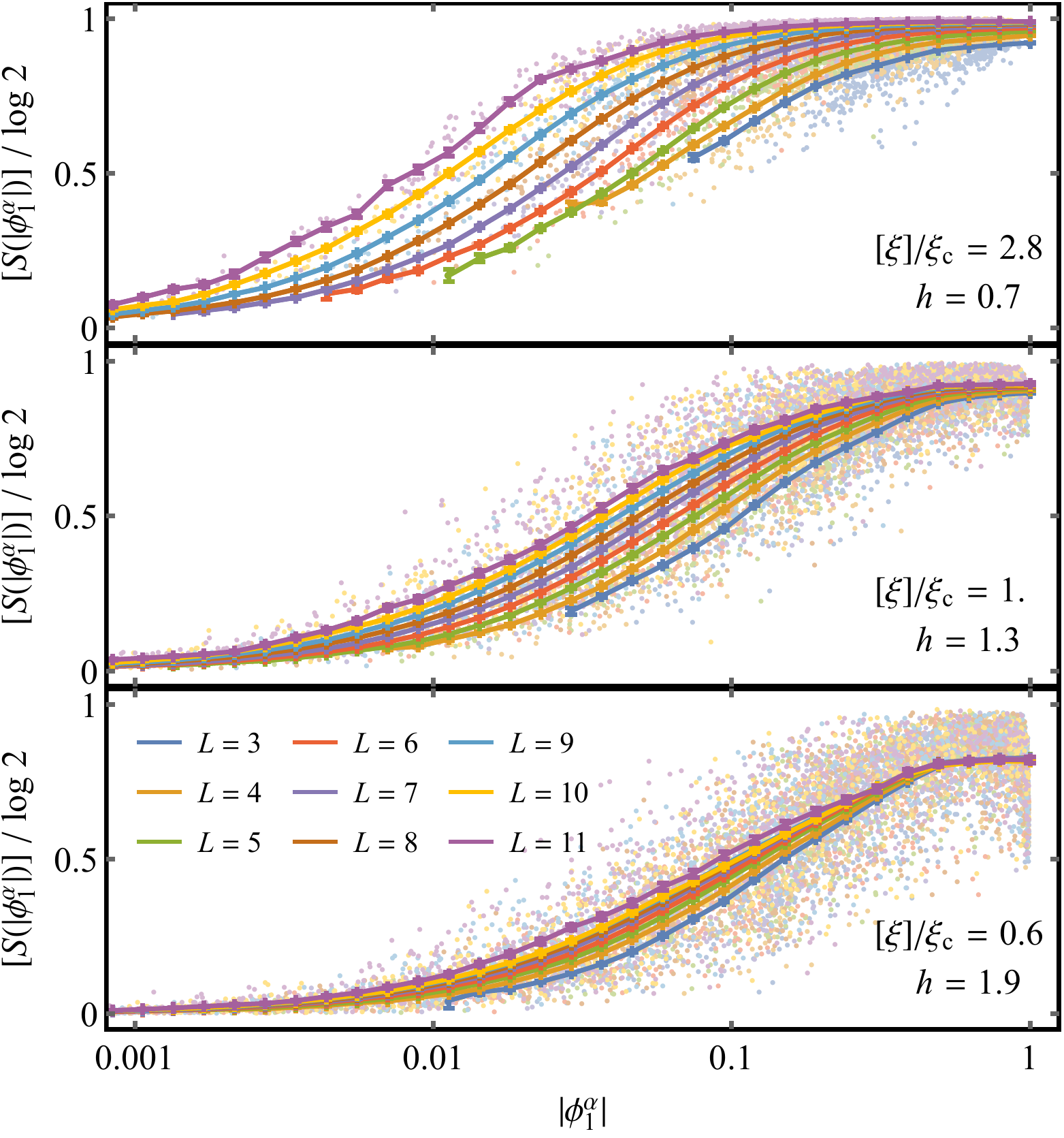}
    \caption{
    \emph{L-bit entanglement entropy in eigenstates as a function of the orbital matrix element:} The eigenstate averaged l-bit entanglement entropy $\bar{S}_\alpha$ is plotted as a function of $|\phi_1^\alpha|$ (points), the orbital overlap which sets the coupling $\tilde{J}_\alpha$ to the thermal seed. Each point corresponds to a sample. The windowed average of $\bar{S}_\alpha$ is additionally shown (solid lines). For $h < \hc$ (top panel), the leftward drift with increasing $L$ (legend) indicates that the matrix element required to thermalize l-bits is exponentially decreasing with $L$. For $h>\hc$ (bottom panel), this drift saturates at sufficiently large $L$. For $h \approx \hc$ (centre panel), a small drift is visible.
    }
    \label{fig:SPhi1}

\end{figure}

Extended avalanche theory (Sec.~\ref{sec:partial_aval}) predicts that the thermal bubble has Hilbert-space dimension $d = d_0 2^{\fA L}$ and density of states $\rho = \rho_0 2^{\fA L}$ for $h$ sufficiently smaller than $\hc$ (so that the typical sample avalanches). Thus, an l-bit hybridises with the bubble if the coupling to the seed ${\tilde J}_\alpha$ exceeds a threshold which is exponentially small in $L$:
\begin{align}
 {\tilde J}_\alpha \rho_0 \sqrt{\frac{ 2^{\fA L}}{d_0 }} >\cS.
 \label{Eq:StrongBubbleCoupling}
\end{align}

 We test Eq.~\eqref{Eq:StrongBubbleCoupling} using the von Neumann entropy of the reduced density matrix obtained by tracing out all l-bits except $\alpha$ in each eigenstate,
\begin{equation}
    S_{\alpha, i} = \tr{\hat{\rho}_{\alpha, i}\log \hat{\rho}_{\alpha, i}}
\end{equation}
where $\hat{\rho}_{\alpha, i} = \ptr{\ket{\psi_i}\bra{\psi_i}}{\bar{\alpha}}$
is the reduced density matrix of the $\alpha$th l-bit in an eigenstate $\ket{\psi_i}$ taken from the middle third of the energy spectrum of the Anderson-seed chain. The eigenstate averaged quantity $\bar{S}_\alpha$ is the defined by averaging $S_{\alpha, i}$ over eigenstates taken from the middle third of the spectrum. Heuristically we expect:
\begin{equation}
     \bar{S}_\alpha= 
    \begin{cases}
    0 & \text{ (localized l-bit)}
    \\
    \log 2 & \text{ (thermal l-bit)}
    \end{cases}
\end{equation}
With each value of $\bar{S}_\alpha$, we also record the orbital overlap of the l-bit onto the first site of the chain $\phi_1^\alpha$. Our data thus consists of a list of pairs $(\bar{S}_\alpha,\phi_1^\alpha)$ from all l-bits $\alpha$ and different samples.

As we seek to measure the average coupling strength necessary to thermalize an l-bit, we further average $\bar{S}_\alpha$ over a window of values of $\phi_1^\alpha$:
\begin{equation}
    [S(\phi')] = \frac{1}{N} \sum_{\phi_1^\alpha \in [\phi'/r,\phi' r]} \bar{S}_\alpha
    \label{eq:Sphiavg}
\end{equation}
where the normalisation factor $N = \sum_{\phi_1^\alpha \in [\phi'/r,\phi' r]} 1$. In the plots, we use $r=1.5$.
 
In Fig~\ref{fig:SPhi1}, we plot $[S(\phi_1^\alpha)]$ as a function of $\phi_1^\alpha$ (solid lines). The points (faded colours) are a random sub-sample of the data prior to window averaging (Eq.~\eqref{eq:Sphiavg}). Data is shown for chain lengths $L = 3 \ldots 10$ (legend inset) for disorder strengths $h = 0.7, 1.3, 1.9$ which are respectively well below, close to, and well above the theoretically predicted avalanche threshold $\hc = 1.37$. The spread on the unaveraged data is due to the sample-to-sample variation in the final size of the thermal bubble. At weak disorder, the leftward drift in the curve $[S(\phi_1^\alpha)]$ with increasing $L$ indicates the exponentially decreasing value of the direct coupling ${\tilde J}_\alpha \sim \phi_1^{\alpha}$ necessary to thermalize an l-bit. This qualitatively confirms the avalanche prediction. This drift disappears at strong disorder where the instability is not present, and thus the thermal bubble is not growing with $L$.

To quantitatively study the drift with $L$ identified in Fig~\ref{fig:SPhi1}, for each $L, h$ we extract the threshold value  $\phi^*$ at which
\begin{align}
[S(\phi^*)] = 0.7 \log 2
\end{align}
For $\phi_1^\alpha \lesssim \phi^*$, the coupling is sufficiently small so that the l-bit is not be fully absorbed into the thermal bubble. Eqs.~\eqref{Eq:StrongBubbleCoupling} and ~\eqref{Eq:fAgeneralform} predict the following behavior for $\phi^*$ as $L \to \infty$:
\begin{equation}
    \phi^* \sim 
    \begin{cases}
    \e^{- \fA L / \xic} \, \text{ with } \fAc \leq \fA \leq 1 & \text{(Avalanches)}
    \\
    \mathrm{constant} & \text{(No instability)}
    \end{cases}
    \label{eq:phistar}
\end{equation}

\begin{figure}
    \centering
    \includegraphics[width=0.95\columnwidth]{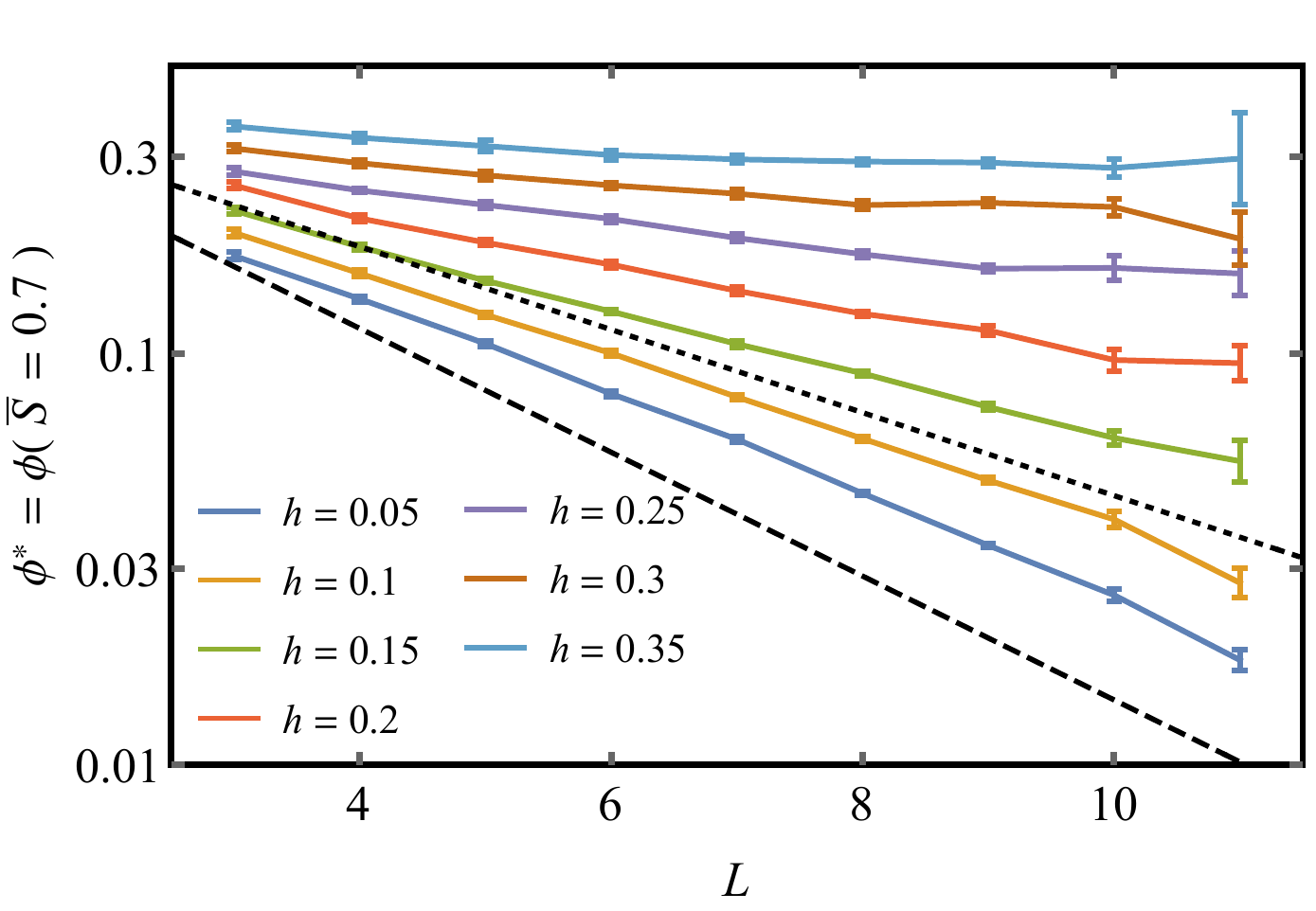}
    \caption{
    \emph{The threshold coupling strength for l-bit thermalization vs $L$:} At small disorder, the avalanche is almost complete and the threshold coupling strength decays as $\phi^* \sim 2^{-L/2}$ (dashed line). For $0<h< \hc = 1.37$, we see that $\phi^* \sim 2^{-a L/2}$ with $a$ continuously decreasing from one (dashed line) to $\fAc$ (dotted line). In contrast, for $h > \hc$, $\phi^*$ saturates to a constant value as $L \to \infty$. 
    }
    \label{fig:PhiL1}

\end{figure}

The numerically extracted values of $\phi^*$ are plotted in Fig.~\ref{fig:PhiL1}. The figure confirms that $\phi^*$ saturates at large $h$ ($h \gg \hc = 1.37$) and exponentially decreases with $L$ at smaller values of $h$ ($h \ll \hc = 1.37$).  
We note that the rate of exponential decrease of $\phi^*$ decreases with $h$ for $h < \hc$, and approaches $\fAc$ near $h = \hc$ (dotted line), as predicted by extended avalanche theory.
We also observe that the saturation value of $\phi^*$ as $L \to \infty$ increases with $h$ above threshold. This is consistent with a finite thermal bubble whose size decreases with increasing disorder strength.

To confirm that the observed thermalization cannot be explained by direct hybridisation with the bare seed, we note that $\phi^* = 0.016$ for $h=0.7$, $L=11$. This value corresponds to a bare matrix element to level spacing ratio of
\begin{equation}
    \frac{{\tilde J}_\alpha \rho_0}{\sqrt{d_0}} = \frac{J \phi^* \rho_0}{2\sqrt{d_0}} = 0.011 \ll \cS = 0.31
\end{equation}
which is more than an order of magnitude below $\cS$, the ratio necessary to induce thermalization in the absence of the thermal bubble, and approximately equal to $\cW = 0.01$, where the effect of the bare coupling is typically perturbative.

Near $h = \hc$, it is unclear whether there is a runaway thermalization instability in finite-size numerics.
As discussed in Sec.~\ref{sec:threshold}, the stopping probability is significant for $l_{\mathrm{dir}} < \mathrm{max}(l^\star, l_\mathrm{FS})$, and the probability of starting an avalanche rapidly goes to zero as $h \to \hc^-$. 
Moreover, the fraction of l-bits absorbed into the thermal bubble will appear to be non-zero in a crossover region around $h = \hc$ at finite $L$ (see Fig.~\ref{fig:Scales}).
As the $L$ range is limited, we do not attempt finite-size scaling here, and leave a detailed numerical study of the threshold to future work.

\subsection{Exponential enhancement of the bath}
\label{sec:v}

\begin{figure}
    \centering
    \includegraphics[width=0.95\columnwidth]{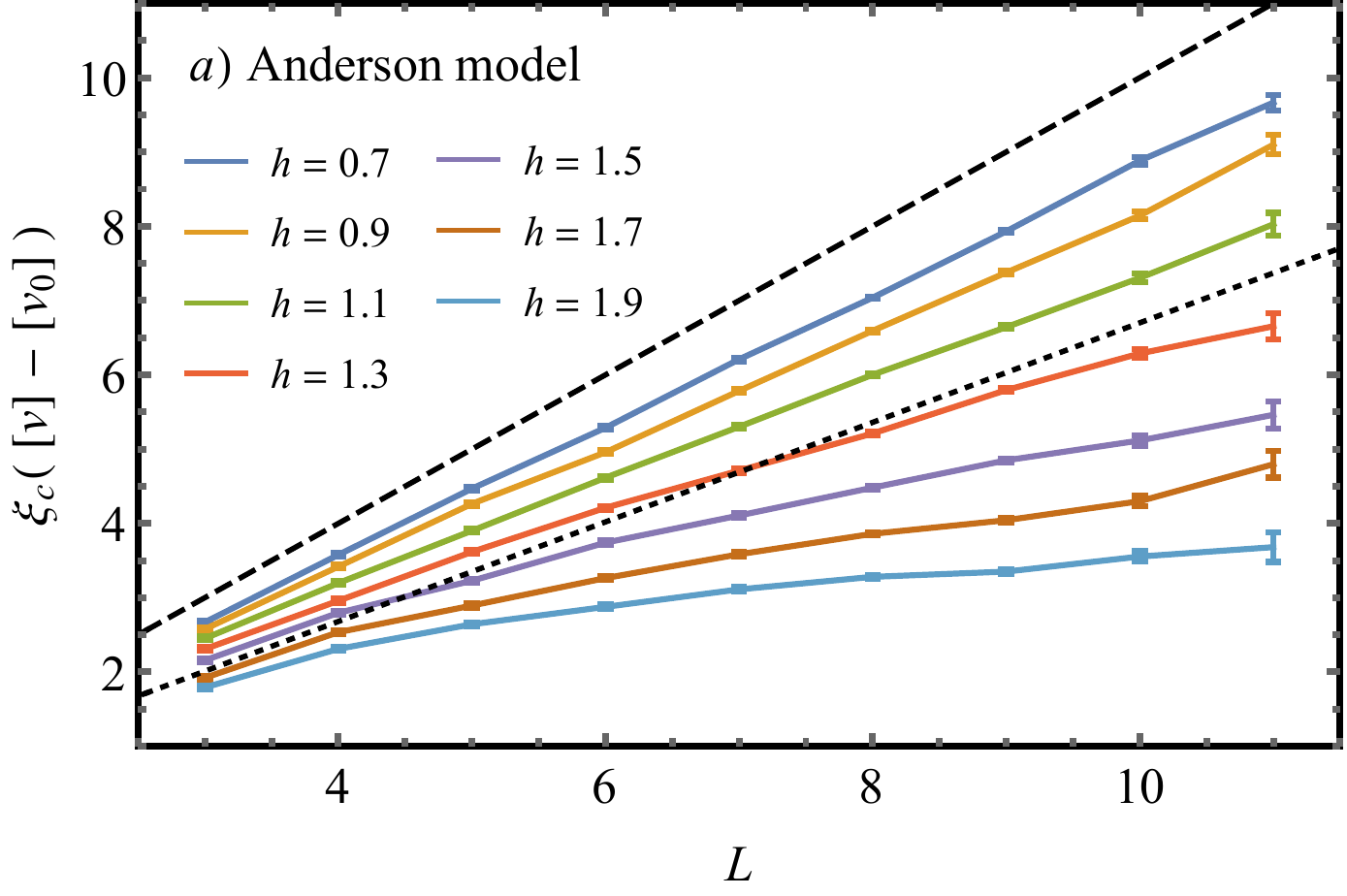}
    \includegraphics[width=0.95\columnwidth]{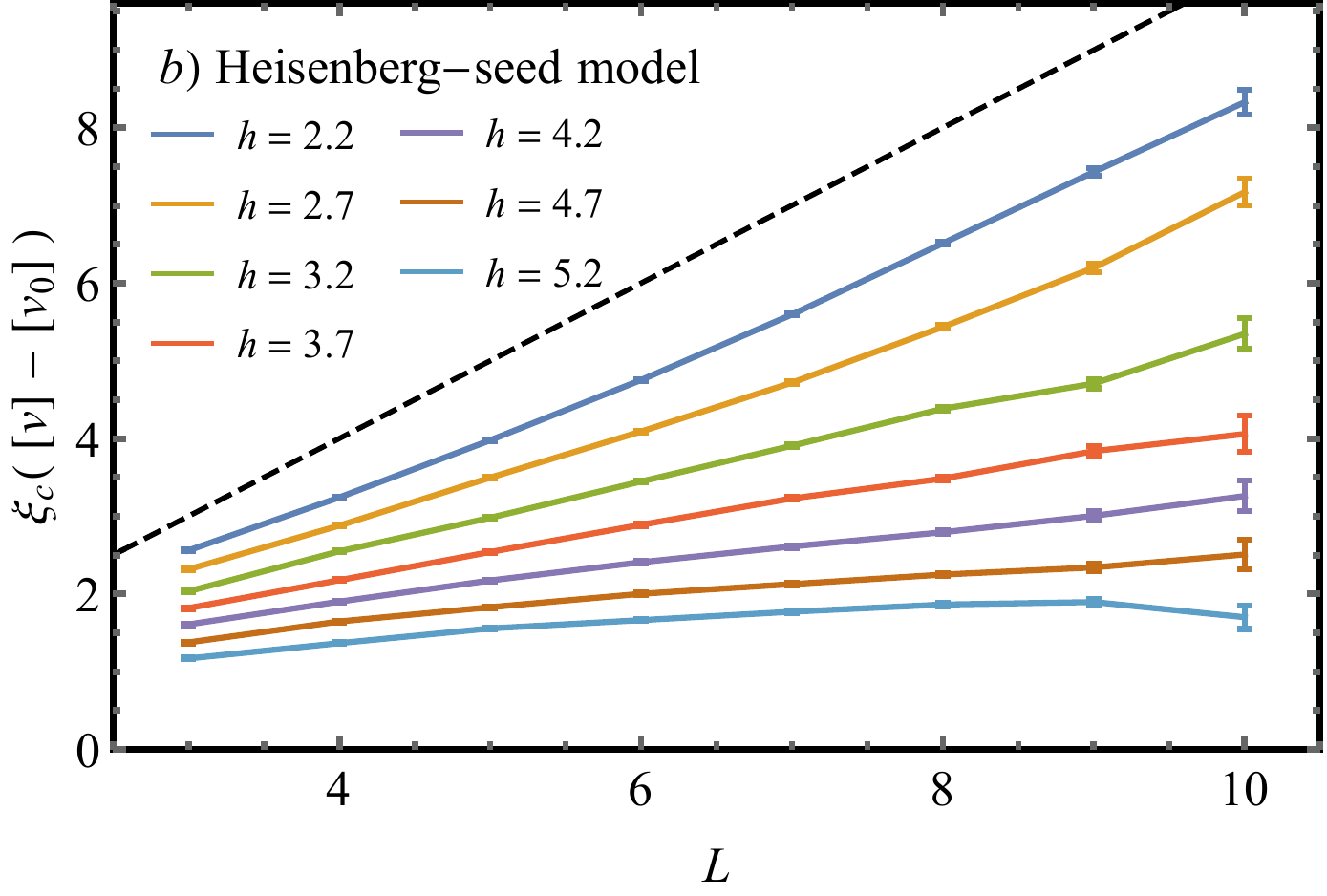}
    \caption{\emph{Scaling of $[v]$ with L:} Upper panel: The matrix element to level spacing ratio $[v] \sim \fbig L /\xic$ is plotted as a function of chain length $L$ in the Anderson-seed model. Data is shown for various disorder strengths (legend inset) above and below the transition $\hc = 1.37$. For a fully thermalizing model $[v] \sim L /\xic$ (dashed), at threshold in the Anderson-seed model $[v] \sim \fAc L /\xic$ (dotted), whereas below threshold $[v]$ saturates. These different behaviours are reproduced for $h$ far above and far below the transition. Lower panel: Analogous data is shown for the Heisenberg-seed model in the vicinity of the MBL transition ($\hc^\mathrm{MBL} \approx 3.7$). The crossover from growing to saturating behaviour is consistent with this value.
    }
    \label{fig:vL}

\end{figure}

Extracting $\phi^*$ requires direct access to the l-bits in the chain. The method in Sec.~\ref{Sec:Sphistar} thus cannot be easily applied to interacting chains. In this section, we introduce a new statistic which measures the thermal bubble size and is easily generalisable to interacting chains.

Consider the full Hamiltonian $H$ (Eq.~\eqref{eq:fullH}) with eigenstates $\ket{\psi_i}$ and eigen-energies $E_i$. Let us introduce a hypothetical probe spin with fixed energy splitting $h_\mathrm{P}$. This spin is coupled to the thermal seed by the same operators as the disordered chain, so that the part of the Hamiltonian involving the probe spin is:
\begin{equation}
    H_{\mathrm{P}} = \tfrac{1}{2} \left(\sigma_{\mathrm{T}}^+ \otimes \sigma_{\mathrm{P}}^- + \sigma_{\mathrm{T}}^- \otimes \sigma_{\mathrm{P}}^+ +   h_\mathrm{P} \bm{1} \otimes \sigma_{\mathrm{P}}^z \right).
\end{equation}
The two states $\ket{\psi_i}\otimes\ket{\uparrow}$ and $\ket{\psi_j}\otimes\ket{\downarrow}$ are related by the matrix element
\begin{equation}
\begin{aligned}
    V_{ij}' &= \tfrac{1}{2} \bra{\psi_i} \sigma_{\mathrm{T}}^- \ket{\psi_j} \bra{\uparrow} \sigma_{\mathrm{P}}^+ \ket{\downarrow}
\end{aligned}
\end{equation}
and separated by the level spacing 
\begin{equation}
    \redd{\delta_{ij}'} = E_i - E_j + h_\mathrm{P}.
\end{equation}
By the arguments in Sec.~\ref{sec:goodthermalseed}, the two states hybridise if $V_{ij}'$ is much larger than $\redd{\delta_{ij}'}$. Thus, the state $\ket{\psi_i}\otimes\ket{\uparrow}$ will hybridise with at least one other state if the quantity 
\begin{equation}
    v_i = \max_j \log \left| \frac{V_{ij}'}{\redd{\delta_{ij}'}} \right|
    \label{eq:vi}
\end{equation}
is sufficiently large. We take the logarithm in Eq.~\eqref{eq:vi}, to capture the typical value as the ratio $V_{ij}'/\redd{\delta_{ij}'}$ is broadly distributed. Finally, we calculate the mean $[v]$ by averaging over the ensemble of Hamiltonians $H$ and over states $i$ from the middle third of the spectrum.

What does $[v]$ measure? We argue that as
\begin{equation}
    \max_j \frac{V_{ij}'}{\redd{\delta_{ij}'}} \sim \frac{\rho}{\sqrt{d}} \sim \e^{\fA L / \xic},
\end{equation}
the statistic $[v]$ thus probes the number of l-bits absorbed into the thermal bubble
\begin{equation}
    [v] = \frac{\fA L}{\xic} + [v_0].
    \label{Eq:MeanVScaling}
\end{equation}
Here $[v_0]$ is the value of this statistic in the absence of the coupling between the thermal seed and the disordered chain (i.e. for $V = 0$).

Suppose first that the full Hamiltonian $H$ satisfies the ETH so that $\fA=1$. ETH implies that $\log| V_{ij}'/\redd{\delta_{ij}'}| \sim \tfrac12 L \log 2$ is maximal for $j$ close to $i$. Consequently, $[v]$ approaches $\tfrac12 L \log 2 = L/\xic$ as $L \to \infty$. Next, suppose that the full Hamiltonian is many-body localized so that $\fA=0$. Here, $\log |V_{ij}'/\redd{\delta_{ij}'}|$ is largest for states $j$ at finite energy above/below $i$, so that $[v]$ approaches an $O(1)$ value as $L \to \infty$. Finally, in a delocalized non-ergodic phase with $0<\fA<1$, $\log |V_{ij}'/\redd{\delta_{ij}'}|$ is maximal when the state $j$ only involves reconfigurations of the thermal bubble. Such states are typically separated by exponentially many (in $L$) intervening states. These intervening states differ in the state of the thermal bubble as well as the localized l-bits%~\footnote{The matrix elements between nearby states are typically exponentially small in $L^2$. Consequently, the $\mathcal{G}$ statistic studied by Ref.~\cite{serbyn2015criterion} will not detect partial avalanches}
. Assuming that the thermal bubble satisfies the ETH, we obtain Eq.~\eqref{Eq:MeanVScaling}. 

\red{The quantity $\exp(v_i)$ is identical to the quantity $\mathcal{G}_i$ defined in Ref.~\cite{thiery2018many}. 
%Specifically, $\mathcal{G}_i = \exp(v_i)$.
However, it has not been numerically studied before.
% if the probe spin field $h_\mathrm{P} = 0$. Quantitatively, the tail of the distribution of $\mathcal{G}_i$ is different when $h_\mathrm{P}= 0$ as compared to when $|h_\mathrm{P}| > 0$ due to the effects of energy level repulsion. Nevertheless, the typical value of  $\mathcal{G}_i$ exhibits the same scaling with $L$ in the ETH, MBL and the delocalized non-ergodic phases in the two cases.
The quantity $v_i$ is however distinct from the $\mathcal{G}$-statistic introduced in Ref.~\cite{serbyn2015criterion} because the latter parameter is defined with $j=i+1$ rather than with $\max_j$ in Eq.~\eqref{eq:vi}. As the matrix elements between nearby states of a delocalized non-ergodic phase are typically exponentially small in $L^2$, the $\mathcal{G}$-statistic studied in Ref.~\cite{serbyn2015criterion} will not detect partial avalanches.}

Fig.~\ref{fig:vL}a shows $[v] - [v_0]$ as a function of $L$ for the Anderson-seed model. Here we take $h_\mathrm{P} = 2\pi \WB/d_0$. At sufficiently small disorder strengths ($h < \hc$), typical samples are predicted to avalanche, and the lines $[v] - [v_0]= \fAc L/\xic$ (black dashed line), and $[v]- [v_0] = L/\xic$ (black dotted line) respectively provide the lower and upper bounds for the growth rates of the thermal bubbles. Indeed, the curves at $h= 0.7, 0.9$ and $1.1$ appear to grow linearly with $L$ with slopes between $\fAc/\xic$ and $1/\xic$. Above threshold ($h > \hc$), there is no runaway thermalization instability, and we observe that $[v]$ indeed saturates. The crossover between linearly increasing and saturating behavior is approximately at $h = \hc$. 

\begin{figure}
    \centering
    \includegraphics[width=0.95\columnwidth]{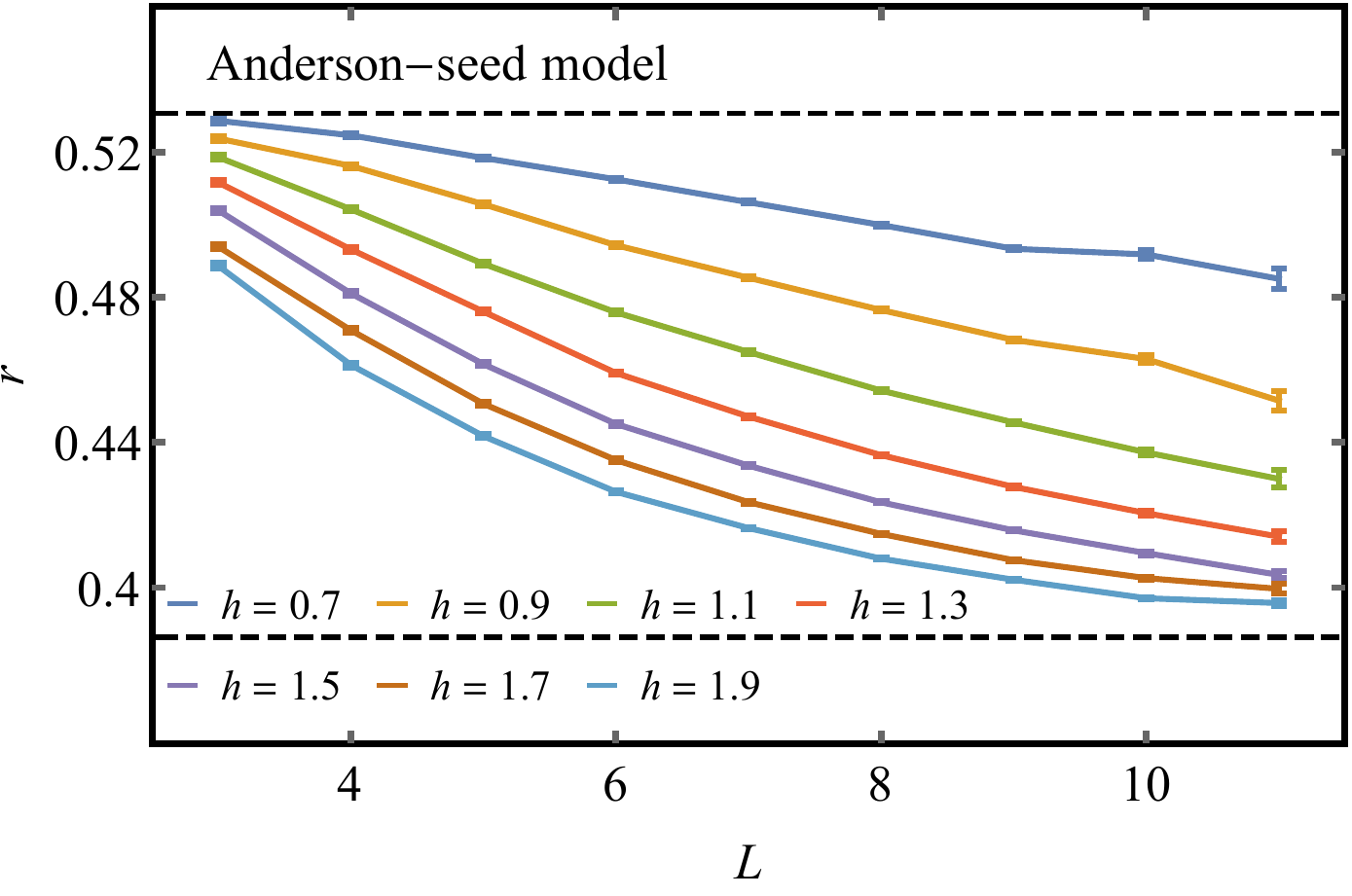}
    \caption{\emph{Variation of level spacing ratio $\bar{r}$ with L in the Anderson-seed model:} the mean level spacing ratio is plotted over the middle third of the spectrum for various disorder strength in the Anderson-seed model. The thermal (GOE) value $\bar{r} = 0.531 \ldots$ and the localized (Poisson) values $\bar{r} = 0.386\ldots$ are shown (horizontal dashed lines). For all disorder strengths (legend inset) the flow is towards the localized value, with no indication of flow reversal. This is due to the coexistence of l-bits with the thermal bubble even below avalanche threshold.
    }
    \label{fig:rL}

\end{figure}

In Fig~\ref{fig:rL} we show spectral statistics for the Anderson-seed model with the same parameters as Fig.~\ref{fig:vL}a . Specifically, we plot the mean level spacing ratio $\bar{r} = [\min(r_i,r_i^{-1})]$ where $r_i = (E_{i+1}-E_i)/(E_i-E_{i-1})$ and averaging is performed over the middle third of spectrum, and over the ensemble of Hamiltonians. $\bar{r}$ measures level repulsion and takes the values $\bar{r} = 0.531\ldots$ in an ergodic phase, and $\bar{r} = 0.386$ in a localized phase~\cite{oganesyan2007localization,atas2013distribution}. We see that, in the Anderson-seed model, for all disorder strengths, $\bar{r}$ flows towards the localized value with increasing chain length $L$. Level repulsion is absent in this model as even below threshold, where the model avalanches, there are many localized l-bits which coexist with the thermal bubble.

\subsection{Heisenberg}
\red{We now turn to the numerical study of the Heisenberg-seed model using the $v$-statistic introduced in Sec.~\ref{sec:v}.}
\label{Sec:NumericsHeisenberg}
\subsubsection{Parameters}

We set $J=1$, $\Delta = 1$, $\WB = 1.6$ and $d=32$. We collect data in the vicinity of the MBL-transition in the random Heisenberg model at $h = h^\mathrm{MBL}_\mathrm{c}  \approx 3.7$~\cite{vznidarivc2008many,pal2010many,devakul2015early,luitz2015many}.

These parameters ensure the bandwidth condition is satisfied
\begin{equation}
    4 \WB = 6.4 > \sqrt{\tfrac32 J^2+ \tfrac34\Delta^2 J^2 + h^2} = 3.99\ldots
\end{equation}
as is the condition to thermalize the first l-bit
\begin{equation}
    \frac{\tilde{J}_1 \rho_0}{ \sqrt{d}} = 0.33 > \cS = 0.31.
\end{equation}
Here we have used that $J_1 = \tfrac12 J \psi_1^1$, $\psi_1^1 \approx 1/\sqrt{\xi}$, at threshold $\xi = \xic$, and the GOE density of states at maximum entropy $\rho_0 = d_0/(\pi \WB)$. We numerically check that our choice of parameters leads to strong hybridisation between the seed and the first physical bit (data not shown).

The length condition, Eq.~\eqref{eq:size_cond}, $L > 10.9$ is model independent and the same as in the Anderson-seed model.

\subsubsection{Exponential enhancement of the bath}

Fig.~\ref{fig:vL}(b) shows the ensemble and spectrally averaged mean $[v]$ (Eq.~\eqref{eq:vi}) as a function of the chain length $L$ for varying disorder strengths $h$ (legend inset). The dashed line shows the theoretical upper bound of $[v]= L/\xic$, when the entire chain satisfies the ETH. At large disorder, we see that $[v]$ saturates, indicating that the interacting chain is stably localized. At low disorder, $[v]$ grows with $L$; the slope shows a slight increase with $L$ towards the ETH value (dashed line) for $h < \hc^\mathrm{MBL} \approx 3.7$. \red{Our finite-size numerics is thus consistent with a delocalized ETH phase with a thermal bubble size $\sim L$.} Remarkably, the value of $h$ at which the scaling behavior of $[v]$ changes is close to the critical value reported in previous studies $\hc^\mathrm{MBL} \approx 3.7$~\cite{vznidarivc2008many,pal2010many,devakul2015early,luitz2015many}, and does not seem significantly altered by the presence of the seed. 

If avalanches underlie the transition out of the MBL phase, we expect the inclusion of the thermal seed to enhance the delocalization tendency at finite chain lengths, and thus shift the location of the transition to larger $h$. However, we do not see clear evidence of this shift at accessible system sizes.

\section{Discussion}
\label{sec:conclusions}
We have extended and tested the theory of quantum avalanches in several respects.
\begin{enumerate}
    \item Assuming that a large enough thermal bubble forms, a single seed partially thermalizes a system with a distribution of l-bit localization lengths with $[\xi]<\xic$. The resulting avalanched phase violates the ETH and is delocalized but non-ergodic.
    \item Thermal bubbles need to reach a certain size in a typical sample before they cause runway thermalization. The required bubble size diverges near the avalanche threshold and is set by the larger of the two lengths scales $l^\star$ and $l_\mathrm{FS}$ (Fig.~\ref{Fig:CriticalFan}).
    \item For the Anderson chain coupled to a single thermal seed, we derived the critical disorder strength $\hc$ below which the chain partially avalanches, the fraction of l-bits absorbed into the thermal bubble for $h<\hc$, and the system parameters to typically observe avalanches.
    \item We introduced a new dimensionless measure $[v]$ of the effective size of the thermal bubble that the seed is a part of, and numerically confirmed that the Anderson-seed model partially avalanches. 
\end{enumerate}
We describe a few broader implications of our results below.

The thermal bubble typically absorbs $\sim l_\mathrm{dir}$ l-bits as $h \to \hc^-$. However, rarely, it grows past the length scale $l_\mathrm{asy} \sim \textrm{max}(l_\mathrm{FS}, l^\star)$, and absorbs $\sim \fAc L$ l-bits. On blocking the chain into regions of size $l_\mathrm{dir}$, the probability of the bubble growing to size $l_\mathrm{asy}$ is seen to be exponentially small in the ratio $l_\mathrm{asy}/l_\mathrm{dir}$. We therefore predict that the mean fraction of l-bits in the thermal bubble as $h \to \hc^-$ is:
\begin{align}
    \fbig(h, l_\mathrm{dir}) \sim \fAc \e^{- |h-\hc|^{-2}/l_\mathrm{dir}} \label{Eq:fbigturnon}
\end{align}
Thus, the growth of the probability in region II of Fig.~\ref{Fig:CriticalFan} with $|h-\hc|$ \emph{is slower than any power law}. This low probability may explain the absence of avalanches reported in Ref.~\cite{goihl2019exploration}, as we discuss further below. 

\red{Avalanches are suppressed near the threshold even if the localization length is unique (as is believed to be the case in MBL systems).
A thermal bubble of size $l$ now has significant stopping probability for $l \ll l^\star$ because of the broad distribution of the couplings between the l-bits and the seed~\cite{varma2019length,gopalakrishnan2019instability}.
Taking the log-couplings to be independently normally distributed with mean $[\log\tilde{J}_\alpha] \sim \alpha$ and variance $\mathrm{Var}(\log\tilde{J}_\alpha) \sim \alpha$, we obtain the following expression following the steps in Sec.~\ref{sec:threshold}:
\begin{equation}
    p_\mathrm{stop}(l) \sim \e^{ - (l/l_{\mathrm{asy}})^{2} }
    \label{Eq:PstopUnique}
\end{equation}
with $l_\mathrm{asy} \sim \textrm{max}(l_\mathrm{FS}, l^\star)$. Although the exponent controlling the divergence of $l^\star \sim |h- h_c|^{-3/2}$ is larger than that in the Anderson-seed case, the asymptotic behavior of $l_\mathrm{asy}$ sufficiently close to $h=\hc$ is unaffected as $l_\mathrm{asy} \sim l_\mathrm{FS} \sim |h-\hc|^{-2}$. Following the discussion in the previous paragraph, the growth of the probability with $|h-\hc|$ in region II of Fig.~\ref{Fig:CriticalFan} \emph{is slower than any power law} in this case as well.}

The suppression of avalanches near threshold explains a number of contradictory results in the literature. Ref.~\cite{luitz2017small} took the coupling between the l-bit $\alpha$ and the seed to be equal to $J \exp(-\alpha/\xi)$ and reported an ETH avalanched phase induced by a thermal seed of dimension $d = 8$ in a $L = 12$ chain. By neglecting the broad distribution of the couplings, we believe that Ref.~\cite{luitz2017small} significantly over-estimated the probability of realistic models to avalanche close to threshold~\red{\footnote{
Moreover, unlike in spin chains, the on-site fields and the localization length $\xi$ are independent variables in the DRH toy model. By picking small enough on-site fields at $\xi \approx \xic$, Ref.~\cite{luitz2017small} ensured the effective thermalization of nearby l-bits by relatively small seeds.}.} Ref.~\cite{goihl2019exploration} studied a realistic Heisenberg chain and reported that a thermal seed of dimension $d=8$ has no destabilizing effect on the MBL phase.
This result contradicts Fig.~\ref{Fig:CriticalFan}.
As short Heisenberg chains typically do not generate their own large thermal seeds, Fig.~\ref{Fig:CriticalFan} predicts a shift of the MBL-ETH transition to larger disorder strengths on coupling the chain an external seed. 
The apparent contradiction is explained by the $d=8$ seed, which our analysis suggests is too small to start avalanches in typical samples. 

The transition between the localized and avalanched regime is reminiscent of the infinite randomness transition~\cite{ma1979random,fisher1992random, fisher1995critical, fisher1999phase}.
Specifically, for $h < \hc$, the transition is characterized by two length scales which diverge with different exponents, $l^\star$ and $l_\mathrm{FS}$.
The more divergent length scale controls the universal scaling function of the fraction of absorbed l-bits near the avalanche threshold:
\begin{align}
    \fbig(h, l_\mathrm{dir}) \sim  \mathcal{F} ( \delta h^\nu l_\mathrm{dir})
\end{align}
where $\mathcal{F}$ is specified by Eq.~\eqref{Eq:fbigturnon} and $\delta h = |h-h_c|$.  
For uncorrelated disorder, $l_\mathrm{FS}\sim \delta h ^{-2}$ while $l^\star \sim \delta h ^{-1}$, and thus $\nu =2$. As in the infinite randomness case, hyper-uniform correlations in the disorder can reduce the exponent controlling the divergence of $l_\mathrm{FS}$ while leaving the exponent of $l^\star$ unchanged~\cite{crowley2019quantum}. 
When the disorder is sufficiently hyper-uniform that $l^\star$ is the most divergent length scale, the exponent is thus $\nu =1$. The quasi-periodic potentials used in previous studies provide access to this case~\cite{iyer2013many,khemani2017two,chandran2017localization,crowley2018quasiperiodic,crowley2018critical,zhang2018universal,mace2019many}. 

Finally, we comment on the implications of this work for the MBL transition~\cite{pal2010many,agarwal2015anomalous,luitz2015many,chandran2015finite,vosk2015theory,potter2015universal,devakul2015early,zhang2016many,vznidarivc2016diffusive,dumitrescu2017scaling,thiery2017microscopically,khemani2017critical,khemani2017two,thiery2018many,goremykina2019analytically,dumitrescu2019kosterlitz,schulz2019phenomenology,gopalakrishnan2019instability}. Numerically accessible systems of length $L\sim 10$ near the MBL-ETH transition are likely to be in the intermediate coupling regime. Assuming that typical samples contain small thermal seeds, the effective coupling $\tilde{J}_\alpha$ between the $\alpha$th l-bit and the seed takes values between the weak and strong limits for almost all $\alpha$:
\begin{equation}
    \cW \approx 0.01 \lesssim \frac{\tilde{J}_\alpha \rho_0}{\sqrt{d_0}} \lesssim \cS \approx 0.31
\end{equation}
In this regime, our single l-bit study suggests that l-bits partially hybridise with the seed with significant eigenstate-to-eigenstate variation within a single sample. Remarkably, numerical studies of the MBL transition have reported significant eigenstate-to-eigenstate variation in entanglement entropy and other spectral quantities~\cite{khemani2017two,khemani2017critical}. Interesting directions for future work include the quantitative description of the finite-size numerics in these terms and the modification of phenomenological renormalisation group theories of the MBL-ETH transition~\cite{vosk2015theory,potter2015universal,zhang2016many,dumitrescu2017scaling,thiery2017microscopically,thiery2018many,goremykina2019analytically,dumitrescu2019kosterlitz} to account for the intermediate coupling regime.

\begin{acknowledgments}

During this work the authors became aware of a related forthcoming work by L. Colmenarez, D. Luitz, and W. De Roeck which studies avalanches in MBL systems~\cite{de2019forthcoming}. We are grateful to W. De Roeck for discussing these preliminary results.

We are very grateful to D. Huse and S. Gopalakrishnan for multiple detailed discussions during the course of the project. We are also very thankful to V. Khemani, V. Oganesyan, C.R. Laumann and R. Vasseur for many insightful discussions and comments about the Anderson-seed system. This work was supported by NSF DMR-1752759 and the Sloan Foundation through the Sloan Research Fellowship.
Numerics were performed on the Boston University Shared Computing Cluster with the support of Boston University Research Computing Services. 

\end{acknowledgments}

\bibliography{Aval_bib}

\appendix

\clearpage

\section{Numerical evaluation of the threshold values $\cS$, $\cW$}
\label{app:cScW}

\begin{figure}
    \centering
    \includegraphics[width=0.95\columnwidth]{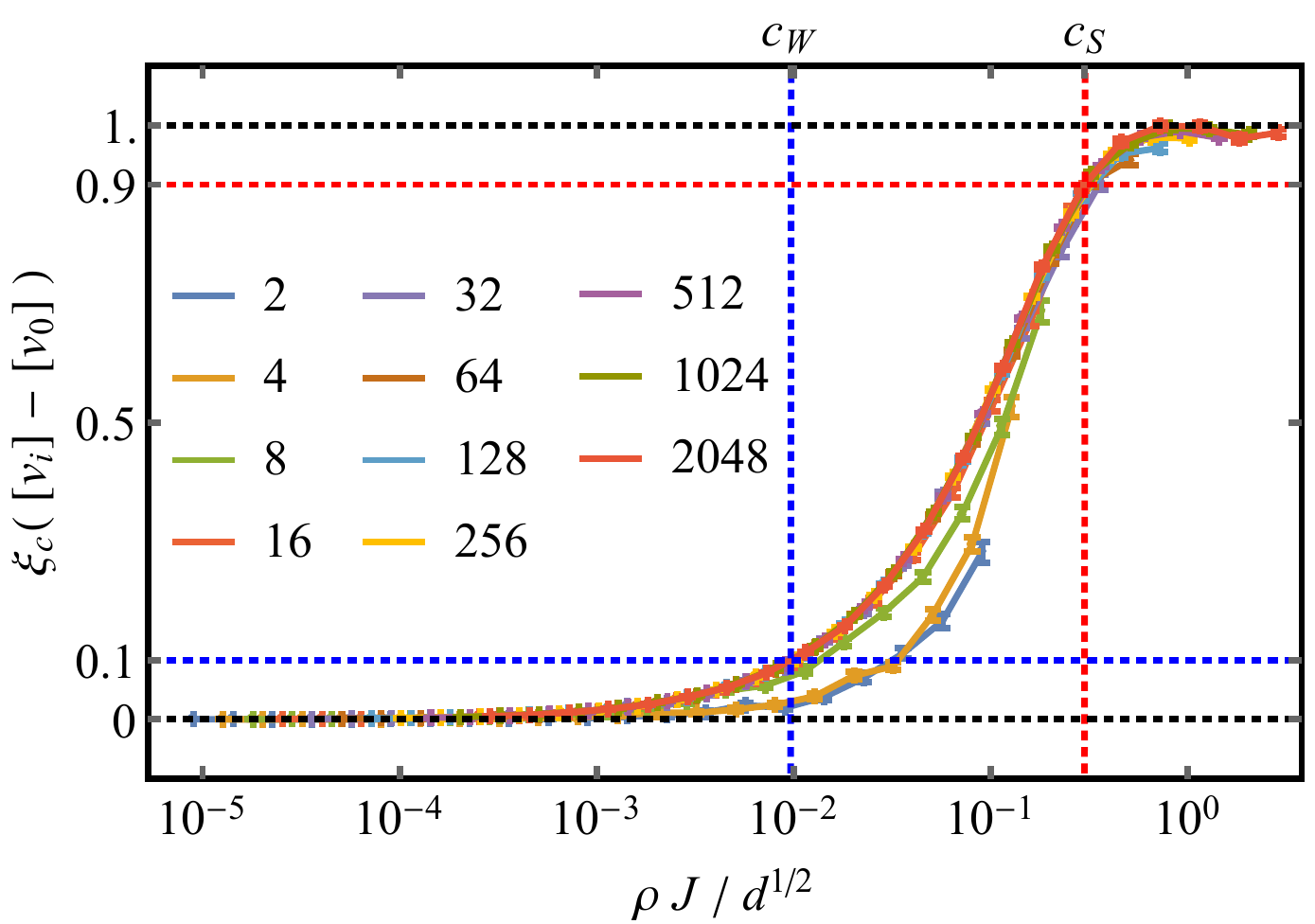}
    \caption{
    \emph{$[v]$ for a GOE matrix coupled to a single spin:} The re-scaled logarithm of the matrix element to level spacing ratio is plotted as a function of the ratio $J \rho/\sqrt{d}$ at different values of $d$ (legend inset) for the model~\eqref{eq:app_model}. This re-scaled quantity has limiting values of zero and one. We identify the values of $[v]$ when it comes within $10\%$ of its limiting values (red/blue horizontal dashed lines). The corresponding threshold values of $J \rho/\sqrt{d}$ are identified as $\cW = 0.01$ and $\cS = 0.31$ (vertical dashed lines).
    }
    \label{suppfig:RdV}

\end{figure}

\begin{figure}
    \centering
    \includegraphics[width=0.95\columnwidth]{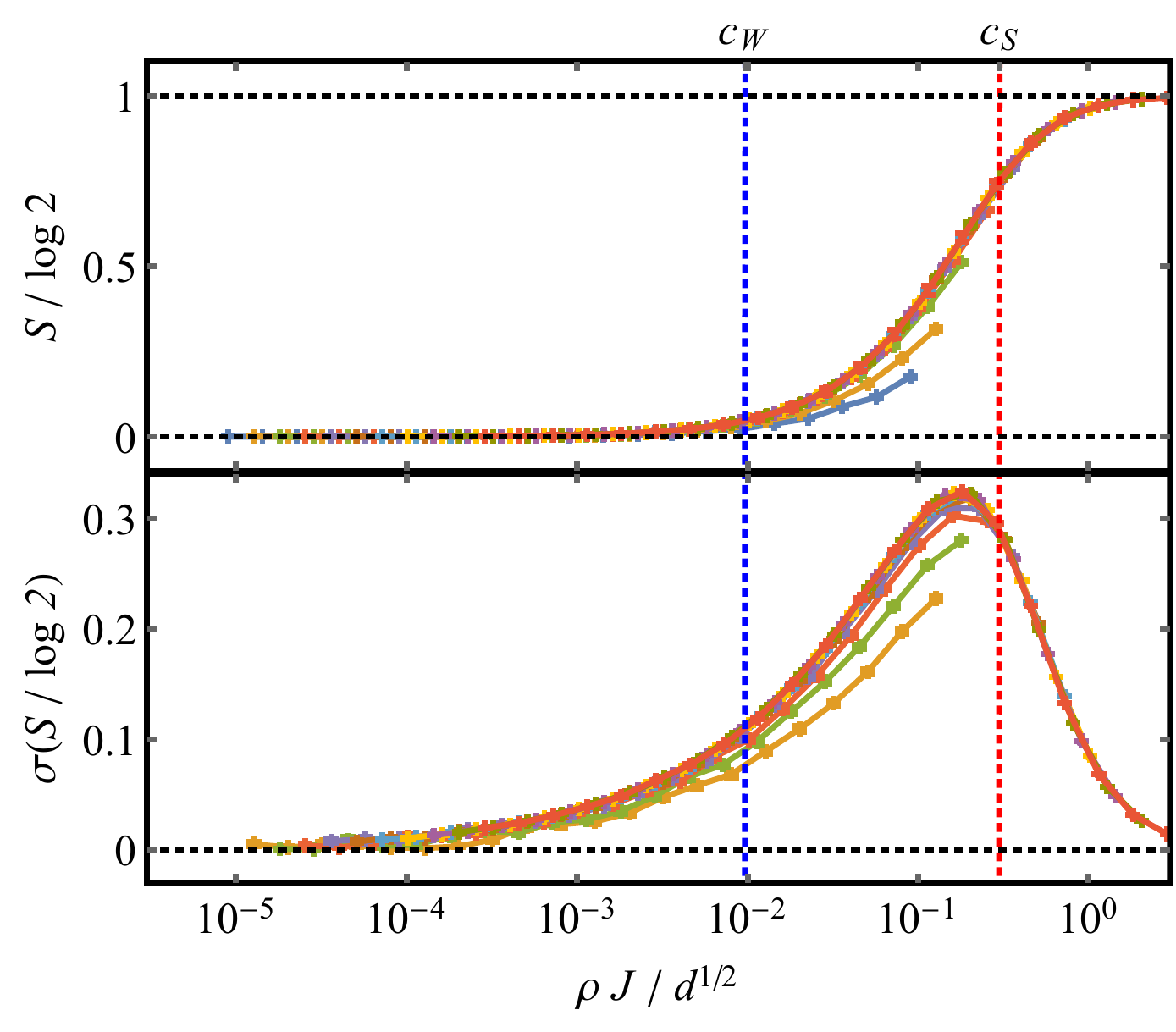}
    \caption{
    \emph{Eigenstate entanglement entropy of a single spin coupled to a GOE matrix:} The mean spin eigenstate entanglement entropy (upper panel) and its intra-sample (i.e. eigenstate-to-eigenstate) standard deviation (lower panel) are plotted as a function of the ratio $J \rho/\sqrt{d}$ for the model~\eqref{eq:app_model}. When $\cW < J \rho/\sqrt{d} < \cS$, the spin is partially hybridized with the GOE matrix, with the hybridization extent showing large eigenstate-to-eigenstate variations. Parameters and legend as in Fig.~\ref{suppfig:RdV}.  
    }
    \label{suppfig:Entropy}

\end{figure}

\begin{figure}
    \centering
    \includegraphics[width=0.95\columnwidth]{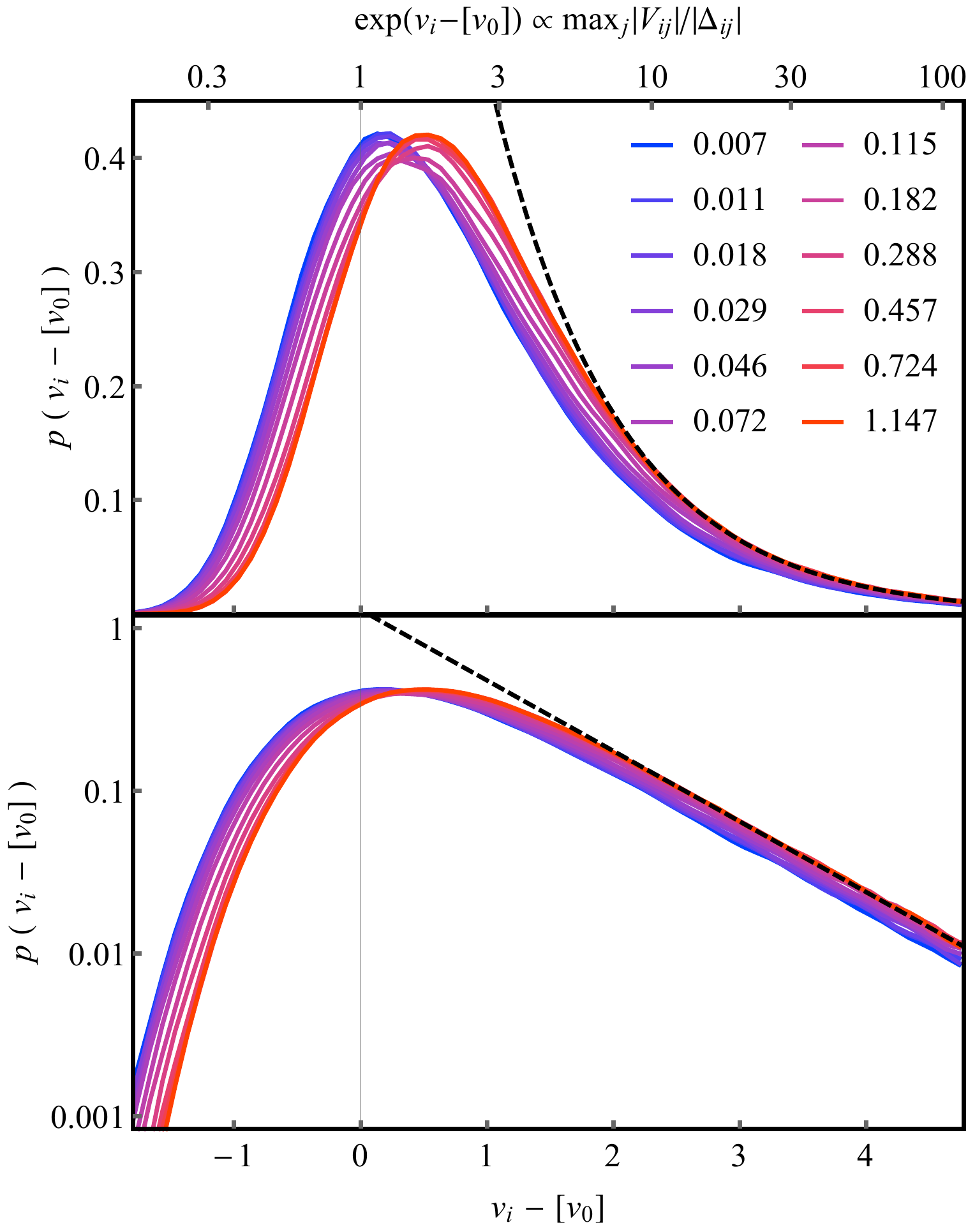}
    \caption{
    \emph{Distributions of $v_i$:} The distributions of $v_i$ for different values of $J \rho / \sqrt{d}$ (legend inset) on a linear scale (upper panel) and a logarithmic scale (lower panel). The widths of the distributions are significantly larger than the drift with increasing $J \rho / \sqrt{d}$. The exponential tail (dashed line) indicates the broadly distributed nature of the ratio $\max_j \left| {V_{ij}'}/{\redd{\delta_{ij}'}} \right|$. Parameters: $d_0 = 2048$, and those given in the caption of Fig.~\ref{suppfig:RdV}.
    }
    \label{suppfig:vDist}

\end{figure}

In this appendix, we numerically determine the constants $\cS$ and $\cW$ introduced in Section~\ref{sec:goodthermalseed} in the main text. 

We extract the threshold values using a set-up consisting of a thermal bubble coupled to a single spin with coupling strength $J$. The thermal bubble's Hamiltonian $R$ is given by a GOE matrix with bandwidth $4 r$, dimension $d$, and density of states at maximum entropy $\rho = d/(\pi r)$. The total Hamiltonian is: 
\begin{equation}
    H = R \otimes \bm{1} + \tfrac{h}{2}\bm{1} \otimes \sigma^z + J (\sigma^+ \otimes \sigma^- + \sigma^- \otimes \sigma^+).
    \label{eq:app_model}
\end{equation}
We use the parameter $v$ (Eq.~\eqref{eq:vi}) defined in Section~\ref{sec:v} to quantify the extent to which the spin hybridizes with the bubble:
\begin{equation}
    [v] = \left[ \max_j \log\left| \frac{V_{ij}'}{\redd{\delta_{ij}'}} \right| \right].
\end{equation}
Here $V_{ij}' = \bra{\psi_i} \sigma^- \otimes \bm{1} \ket{\psi_j}$, $\ket{\psi_i}$ are the eigenstates of $H$, $E_i$ are the eigen-energies of $H$ and the energy splitting in the denominator is given by $\redd{\delta_{ij}'} = E_i - E_j + h_\mathrm{P}$. This corresponds to the matrix element to level spacing ratio felt by a second probe qubit with splitting $h_\mathrm{P}$. The field $h_\mathrm{P}$ on the probe spin is taken to be sufficiently small, so that the density of states of the thermal bubble can be treated as a constant.

The physical regime of interest is when the couplings and fields are greater than the energy level spacing on the bubble, but less than bandwidth of the bubble
\begin{equation}
     \rho^{-1}\ll J, h, h_\mathrm{P} \ll 4 r.
     \label{eq:app_regime}
\end{equation}
In this regime, by the arguments presented in the main text, $[v]$ is a function of the ratio $J \rho/\sqrt d$ only. When $J \rho/\sqrt d$ is sufficiently small, the spin and bubble barely hybridize and $[v]$ is close to the value $[v_0]$ at $J=0$. When $J \rho/\sqrt d$ is sufficiently large, the spin and the bubble strongly hybridise, and $[v]$ approaches its maximal value:
\begin{equation}
    [v] \to [v_0] + \log \sqrt{2} = [v_0] + \frac{1}{\xic}
\end{equation}
We define the following threshold values where $[v]$ comes within $10\%$ of one of these limiting values
\begin{equation}
\begin{aligned}
    {[v]} &< [v_0] + \frac{0.1}{\xic} \quad \text{for} \quad \frac{J \rho}{\sqrt{d}} < \cW,
    \\
    {[v]} &> [v_0] + \frac{0.9}{\xic} \quad \text{for} \quad \frac{J \rho}{\sqrt{d}} > \cS.
\end{aligned}
\end{equation}
In Fig.~\ref{suppfig:RdV}, we plot $[v]$ as a function of $J \rho/\sqrt d$ for the model Eq.~\eqref{eq:app_model} with parameters $h=1$, $h_\mathrm{P} = 1.5$, $r=5$, and $d$ values shown in the inset. The data exhibits good collapse with the exception of small $d$ where the bubble density of states is too low and hence the condition~\eqref{eq:app_regime} is violated. From the collapsed data, we identify the weak and strong thresholds to be $\cW = 0.01$ and $\cS = 0.31$ respectively.

For comparison, in Fig.~\ref{suppfig:Entropy} we plot the mean entanglement entropy of the spin (taken over the middle third of the spectrum), and the intra-sample (eigenstate-to-eigenstate) standard deviation of this quantity for the same parameters as in Fig.~\ref{suppfig:RdV}. We see that:
\begin{align}
    S &\approx 0,  \quad \frac{J \rho}{\sqrt{d}} < \cW \\
    S &\approx \log 2,  \quad \frac{J \rho}{\sqrt{d}} > \cS
\end{align}
For intermediate value of $\frac{J \rho}{\sqrt{d}}$, the spin partially hybridises with the bubble, with large eigenstate-to-eigenstate variation in the extent of hybridisation.

\subsection{Distribution of $v_i$}

Here we discuss the distributions of the quantity $v_i$ (Eq.~\eqref{eq:vi}) across eigenstates within a sample. The underlying quantity 
\begin{equation}
    z = \left|\frac{V_{ij}'}{\redd{\delta_{ij}'}}\right|
\end{equation}
is broadly distributed, as the denominator $\redd{\delta_{ij}'}$ has a probability distribution which remains finite as $\redd{\delta_{ij}'} \to 0$. The probability distribution of $z$ thus decays as a power law at large $z$:
\begin{equation}
    p(z) \sim z^{-2}.
    \label{Eq:PTailZ2}
\end{equation}
For example, the Cauchy distribution, which is the ratio of two normally distributed random variables, satisfies Eq.~\ref{Eq:PTailZ2}.

The definition in Eq.~\eqref{eq:vi} includes a maximum over states $j$ and a logarithm. Taking the maximum over the index $j$ leaves the tail of the $z$-distribution unaltered (Eq.~\eqref{Eq:PTailZ2}). Using the property that the logarithm of a power-law distributed random variable is exponentially distributed: 
\begin{equation}
p(y) \sim y^{-n} \iff p'(\log y ) \sim \e^{- (n-1) \log y} ,
\end{equation}
we conclude that the distribution of $v_i$ decays exponentially at large $v_i$
\begin{equation}
    p(v_i) \sim \e^{-v_i}.
\end{equation}
In Fig.~\ref{suppfig:vDist}, we verify these features. In the upper panel, distributions of $v_i - [v_0]$ are shown for different values of $J \rho/\sqrt{d}$ (values inset in legend). The drift of the distribution to larger $v_i$ with increasing $J \rho/\sqrt{d}$ is precisely the enhancement shown in Fig.~\ref{suppfig:RdV}. The dashed line indicates the expected behavior of $p(v_i) \sim \e^{-v_i}$ at large $v_i$.

A prominent feature of these plots is that the distributions of values across eigenstates is significantly wider than the shift induced by absorbing a single spin. Recalling that $\max_j \left|{V_{ij}'}/{\redd{\delta_{ij}'}}\right| = \e^{v_i}$ sets the extent of hybridisation of the $i$th eigenstate, the width of these distributions is understood as the origin of the significant spread between the constants $\cW$ and $\cS$. Specifically, the width of the distribution entails that for spins (or l-bits) with coupling strengths spanning about two orders of magnitude, there are significant numbers of resonant and non-resonant eigenstates. The peak in the standard deviation of $S$ in the lower panel of Fig.~\ref{suppfig:Entropy} also reflects the wide distribution of the extent of hybridization of the spin across eigenstates.

\section{Various proofs regarding partial avalanches}
\label{app:partial_proofs}

\begin{figure}
    \centering
    \includegraphics[width=0.95\columnwidth]{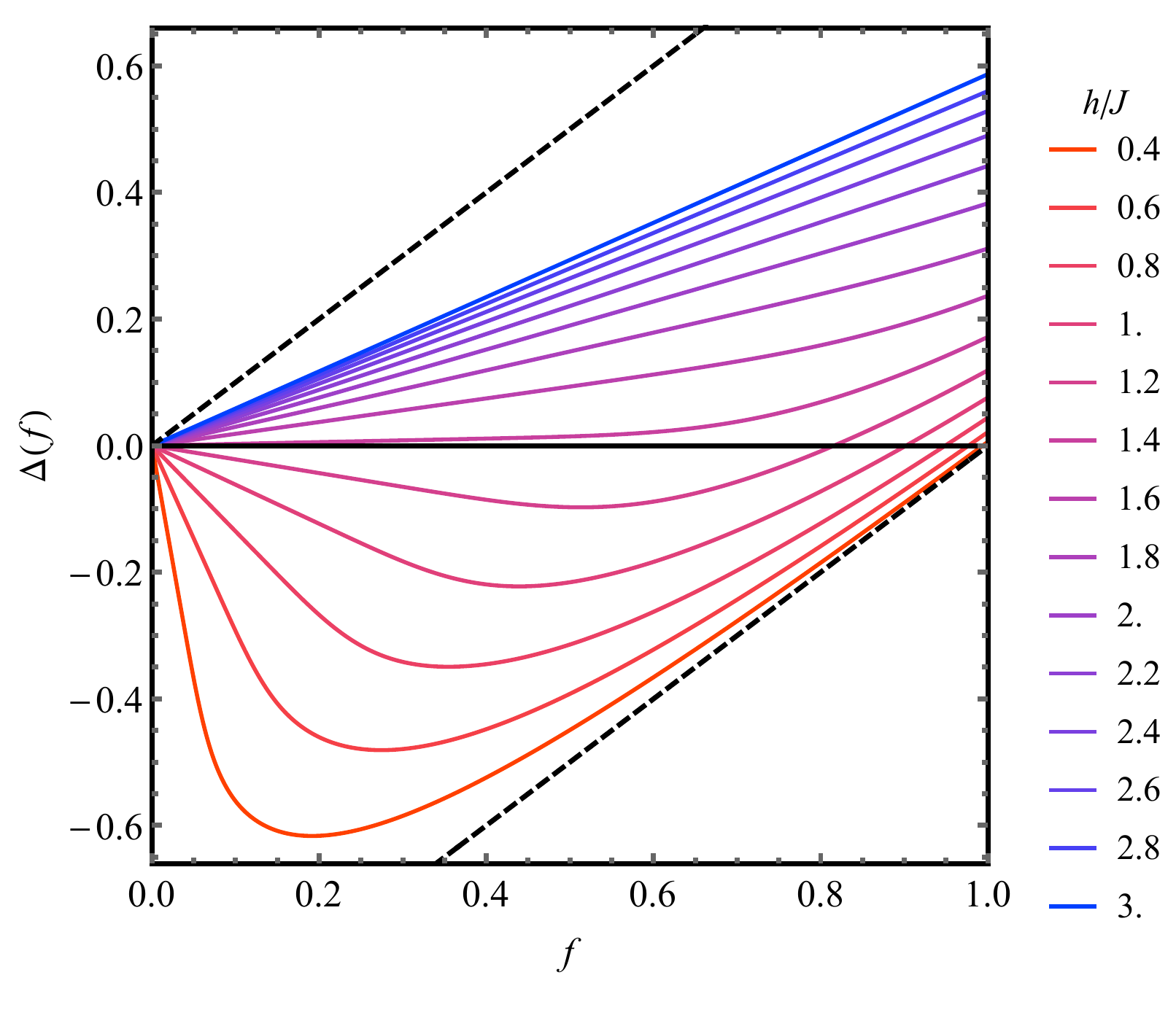}
    \caption{
    \emph{Plots of $\Delta(f)$}: $\Delta(f)$ is plotted for various disorder strengths (see legend). For strong disorder $\Delta(f) \to f$ and for weak disorder $\Delta(f) \to f-1$ (dashed black lines). The solid black line shows $\Delta = 0$. $\Delta(f)$ is asymptotically increasing, convex, and always satisfies $\Delta(0)=0$. These properties are visually clear. From these properties various useful results regarding the roots $\Delta(f)=0$ follow. These forms were computed numerically by exactly diagonalizing the Anderson model at $L=3000$ and $4000$ samples.
    }
    \label{suppfig:Delta}

\end{figure}

In Sec.~\ref{sec:partial_aval}, we introduced the self-consistency condition Eq.~\eqref{Eq:SelfconsistentfA} for the fraction of thermalized l-bits in the disordered chain. In this appendix, we derive several useful results which are stated in the main text. To study Eq.~\eqref{Eq:SelfconsistentfA} in more detail, it is useful to define the quantity
\begin{equation}
    \Delta(f) = f - 1 + \int_0^{\xic / f} d \xi' \left( 1- \frac{ f \xi'}{\xic} \right) p(\xi')
    \label{eq:Delta}
\end{equation}
whose roots $\Delta(f)=0$ are the solutions to the self-consistency equation Eq.~\eqref{Eq:SelfconsistentfA}. $\Delta(f)$ has several useful properties:
\begin{itemize}
    \item {$f = 0$ is always a root}. Specifically, $\lim_{f \to 0^+} \Delta(f) = 0$. 
    \item {$\Delta(f)$ is differentiable for $f \geq 0$}. $\Delta(f)$ has first derivative is given by
    \begin{equation}
        \partial_f \Delta = 1- \int_0^{\xic/f}d\xi'\frac{\xi'}{\xic}p(\xi')
    \end{equation}
    \item {$\Delta(f)$ is convex for $f\geq 0$}: This is seen by further differentiating to obtain
    \begin{equation}
        \partial_f^2\Delta = \frac{\xic p(\xic/f)}{f^3} \geq 0.
    \end{equation}
    \item {$\Delta(f) \geq f-1$}: this is easily seen by noting that the integrand in the second term in the RHS of Eq.~\eqref{eq:Delta} is strictly non-negative over the domain of integration. 
    \item $\Delta(f)$ is continuous provided we make only the assumption that the cumulative distribution function $\int_0^\xi d \xi' p(\xi')$ is continuous. 
\end{itemize}
These properties are all evident in Fig.~\ref{suppfig:Delta}. Using these properties, we prove the following propositions regarding the partial avalanche theory.
\begin{itemize}
    \item \textbf{If $\xia < \xic$ then $\fA = 0$ is the only solution to Eq.~\eqref{Eq:SelfconsistentfA}:} \emph{Proof:} We prove the equivalent claim, that $\Delta(f)$ has exactly one non-negative root, given by $f = 0$. It was previously noted that $f=0$ is always a root. To see that this is the only root for $\xia < \xic$, we note that the first derivative of $\Delta$, evaluated at $f = 0$,
    \begin{align}
        \left.\partial_f \Delta \right|_{f=0} &= \lim_{f \to 0^+} \left( 1 - \int_0^{\xic/f} d\xi' \frac{\xi'}{\xic} p(\xi') \right) \\
        &= 1 - \frac{\xia}{\xic}
        \label{eq:B4}
    \end{align}
    is positive for $\xia < \xic$. As $\Delta(f)$ is convex for $f>0$. Eq.~\eqref{eq:B4} further implies there are no further non-negative roots.
    
    \item \textbf{If $\xia > \xic$ there is exactly one other solution $\fA$ to Eq.~\eqref{Eq:SelfconsistentfA}. } \emph{Proof:} We prove the equivalent claim, that $\Delta(f)$ has exactly two non-negative roots, one at $f = 0$, and one in $f \in (0,1]$. It was previously noted that $f=0$ is always a root. The existence of the second root in the interval $f \in (0,1]$ follows from (i) the first derivative $\left. \partial_f \Delta \right|_{f=0} < 0$, which implies that $\Delta(f)<0$ for sufficiently small $f>0$ (ii) $\Delta(1) \geq 0$, which follows from $\Delta(f) \geq f - 1$ (iii) the continuity of $\Delta(f)$. That there are no further roots follows from the convexity of $\Delta(f)$ for $f>0$.
    
    \item \textbf{If $\xia \neq \xic$, then $\fA \not \in (0, \xic/\xim]$:} \emph{Proof:} We prove the equivalent claim that $\Delta(f)$ has no roots in $(0, \xic/\xim]$ for $\xia \neq \xic$. Assume that $f \in (0, \xic/\xim]$. Then, the support of $p(\xi)$ falls entirely within the domain of integration, and
    \begin{equation}
        \Delta(f) = f\left( 1 - \frac{\xia}{\xic} \right).
    \end{equation}
    Thus, the only root for $f \in (0, \xic/\xim]$ occurs for $\xia = \xic$.
\end{itemize}

\subsection{Evaluation of $p_\mathrm{stop}$ }
\label{app:pstop}

In this section, we derive the result quoted in Sec.~\ref{sec:threshold}, that the probability $p_\mathrm{stop}$ (Eq.~\eqref{eq:pstop_prod}) of a thermal bubble of size $l$ ceasing to grow is exponentially small in $l/l^\star$, where the length scale $l^\star$ diverges as $l^\star \sim |h - \hc|^{-1}$ at the avalanche threshold. We begin by recalling the definition of $p_\mathrm{stop}$
\begin{equation}
\begin{aligned}
    p_\mathrm{stop}(l) &= \prod_{\alpha = l + 1}^{L} \int_0^{\alpha /(\fA l/\xic + l_{\mathrm{W}}/\xia)} d \xi' p(\xi') 
    \\
    &= \exp \left( \sum_{\alpha = l + 1}^{L} \log \int_0^{\alpha /(\fA l/\xic + l_{\mathrm{W}}/\xia)} d \xi' p(\xi') \right)
\end{aligned}
\end{equation}
where we have re-written the product as an exponentiated sum.
As the integrand is zero for $\alpha /(\fA l/\xic + l_{\mathrm{W}}/\xia) > \xi_\mathrm{max}$, we assume a sufficiently long chain $L > l \fA \xim/ \xic$, and replace the upper bound of the sum with infinity.
\begin{equation}
\begin{aligned}
    p_\mathrm{stop}(l) &= \exp \left( \sum_{\alpha = l + 1}^{\infty} \log \int_0^{\alpha /(\fA l/\xic + l_{\mathrm{W}}/\xia)} d \xi' p(\xi') \right)
\end{aligned}
\end{equation}
Next, we replace the sum with an integral. The Riemann sum over $\alpha$ in the above equation is $\mathrm{O}(l^1)$, whereas the error incurred from replacing it with an integral is $\mathrm{O}(l^0)$. Thus, for sufficiently large $l$ we are at liberty to make this replacement. We further neglect the sub-leading in $l$ corrections to the upper bound of the integral over $\xi$. We then obtain
\begin{equation}
\begin{aligned}
    p_\mathrm{stop}(l) &= \exp \left( \int_{l}^{\infty} d \alpha \log \int_0^{  (\alpha \xic)/(\fA l)} d \xi' p(\xi')  + O (l^0)\right)
    \\
    & = \exp \left( \frac{\fA l}{\xic} \int_{\xic/\fA}^{\infty} d x \log \int_0^{x} d \xi' p(\xi') + O (l^0) \right).
\end{aligned}
\end{equation}
In the second line of the equation above we have simply made the replacement $ x = (\xic \alpha)/(\fA l)$. Thus, we see that for $l \gg 1$, $p_\mathrm{stop} \sim \e^{- l / l^\star}$ with $l^\star$ given by 
\begin{equation}
    \frac{1}{l^\star} = -\frac{\fA}{\xic} \int_{\xic/\fA}^{\infty} d x \log \int_0^{x} d \xi' p(\xi').
    \label{eq:lstar}
\end{equation}
This sets the length scale to which the thermal bubble must grow to set off the thermalization instability.

As the avalanche threshold is approached, this length scale diverges. This can be seen by re-writing Eq.~\eqref{eq:lstar} 
\begin{equation}
    \frac{1}{l^\star} = -\frac{\fA}{\xic} \int_{\xic/\fA}^{\xim} d x \log \left( 1-\int_x^{\xim} d \xi' p(\xi') \right).
\end{equation}
Note that (i) $x \in [\xic/\fA,\xim]$, and (ii) $\fA \to \xic / \xim$ as $h \to \hc^-$. Thus, the argument of the logarithm is close to unity, and may be Taylor expanded: 
\begin{equation}
\begin{aligned}
    \frac{1}{l^\star} &= \frac{\fA}{\xic} \int_{\xic/\fA}^{\xim} d x \int_x^{\xim} d \xi' p(\xi') + \ldots
    \\
    & = \frac{\fA}{\xic} \int_{\xic/\fA}^{\xim} d \xi' p(\xi') \left(\xi' - \frac{\xic}{\fA}\right) + \ldots
\end{aligned}
\end{equation}
In the second step, we switched the order of integration, and performed one of the subsequent integrals. The higher order terms may be neglected as they are asymptotically smaller than the leading order terms in the limit of interest, where $1/l^\star \to 0$. Rearranging the above equation and substituting in Eq.~\eqref{Eq:SelfconsistentfA} we then obtain
\begin{equation}
\begin{aligned}
    \frac{1}{l^\star} &= \left( \frac{\fA \xia }{\xic} - 1\right)- \int_0^{\xic/\fA} d \xi' p(\xi') \left(\frac{\fA \xi'}{\xic} - 1\right)
    \\
    & = \left( \frac{\fA \xia }{\xic} - 1\right)- (\fA - 1)
    \\
    & = \fA \left( \frac{\xia }{\xic} - 1\right).
\end{aligned}
\end{equation}
This is the result given in the main text as Eq.~\eqref{eq:critical_lstar}. Sufficiently close to the threshold, we may expand in powers of $h - \hc$, 
\begin{equation}
\begin{aligned}
    \xia & = \xic + \left. \partial_h \xia \right|_{h = \hc}  (h - \hc) + \ldots
    \\
    \fA & = \fAc + \left. \partial_h \fA \right|_{h \to \hc^-} (h - \hc) + \ldots
\end{aligned}
\end{equation}
to obtain
\begin{equation}
    \frac{1}{l^\star} =\frac{ \fAc}{\xic} \left. \frac{\partial \xia}{\partial h} \right|_{h =\hc} (h - \hc) + \mathrm{O}(h-\hc)^2
\end{equation}

\section{Numerical evaluation of the Anderson critical field $\hc$ and threshold thermal fraction $\fAc$}
\label{app:anderson_analysis}

\begin{figure}
    \centering
    \includegraphics[width=0.95\columnwidth]{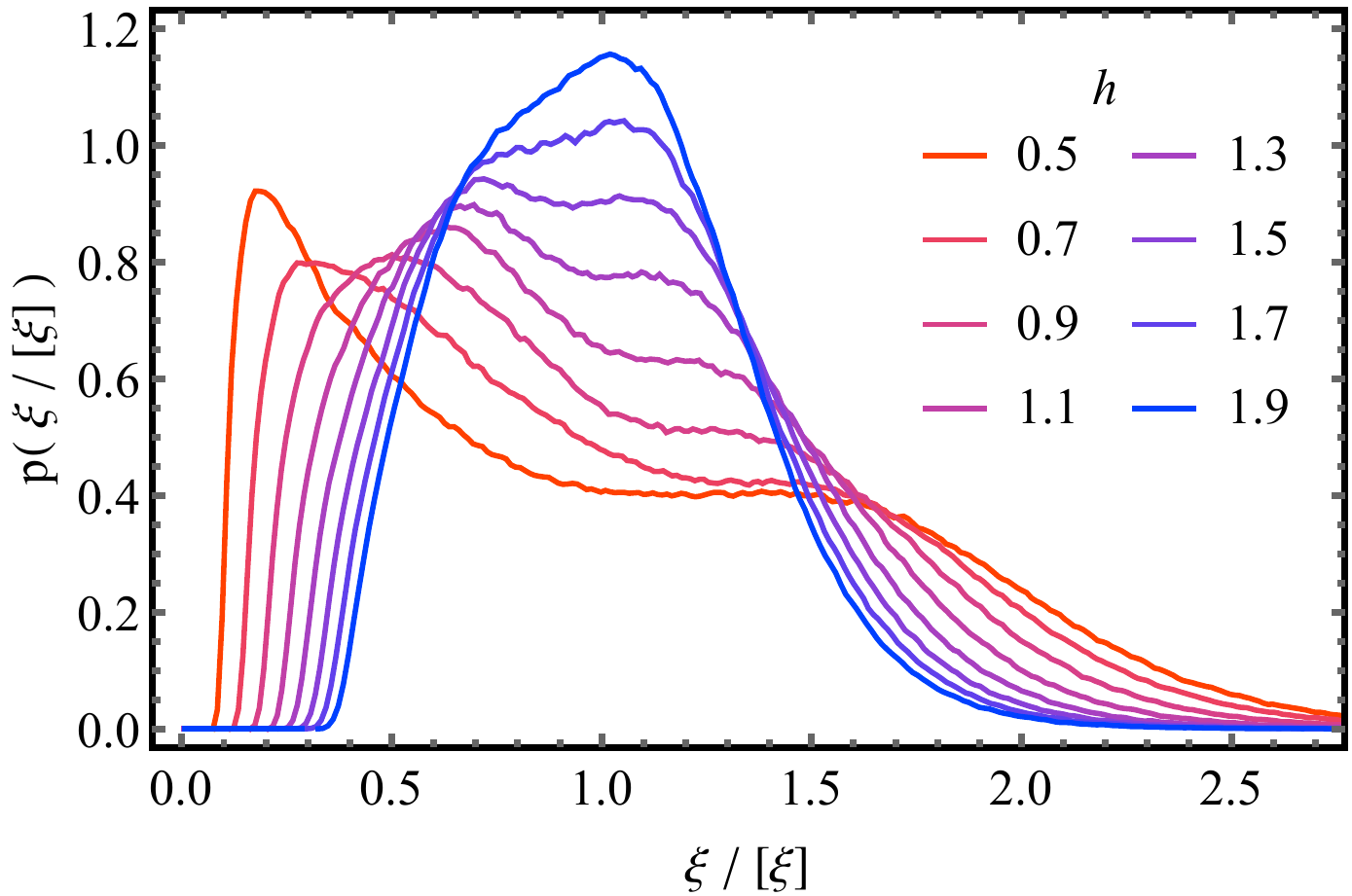}
    \caption{
    \emph{The distribution of re-scaled localization lengths $\xi/\xia$ in the Anderson model:} At large disorder strengths, the distribution is narrow and peaked about one. However, with decreasing disorder, the distribution develops a double peak, with the taller peak moving to smaller values of $\xi/\xia$. The distributions were computed numerically by exactly diagonalizing the Anderson model at $L=3000$ and $4000$ samples.
    }
    \label{suppfig:pxi}

\end{figure}

\begin{figure}
    \centering
    \includegraphics[width=0.95\columnwidth]{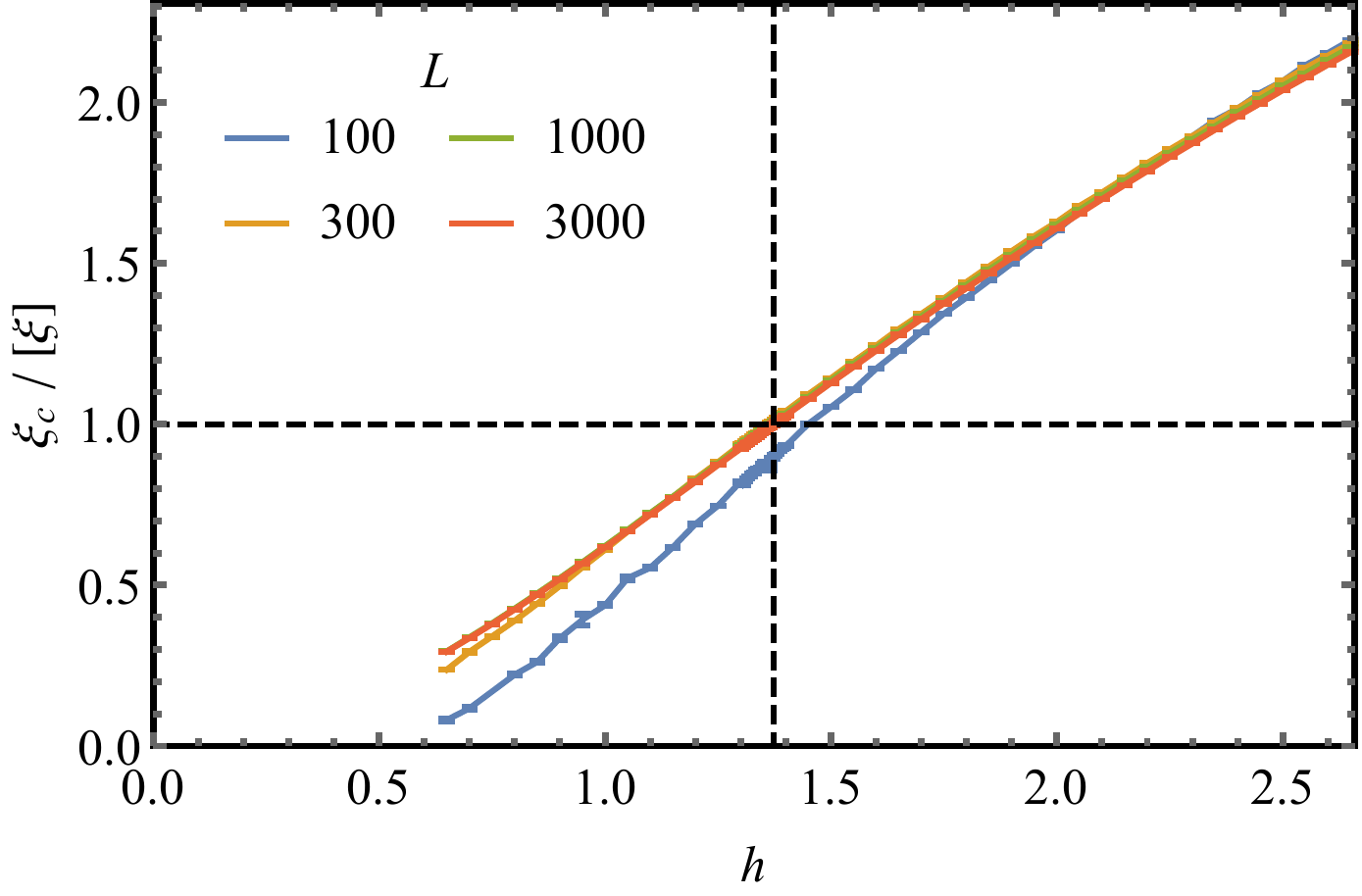}
    \caption{
    \emph{Extracting $\hc$ from the avalanche condition:} At the avalanche threshold, $\xia = \xic$. The plot shows the $\xic/\xia$ as a function of the disorder strength for different chain lengths (legend insets). We find $\hc = 1.37 \ldots$ (vertical dashed line). Data for $4000$ samples.
    }
    \label{suppfig:hc}

\end{figure}

\begin{figure}
    \centering
    \includegraphics[width=0.95\columnwidth]{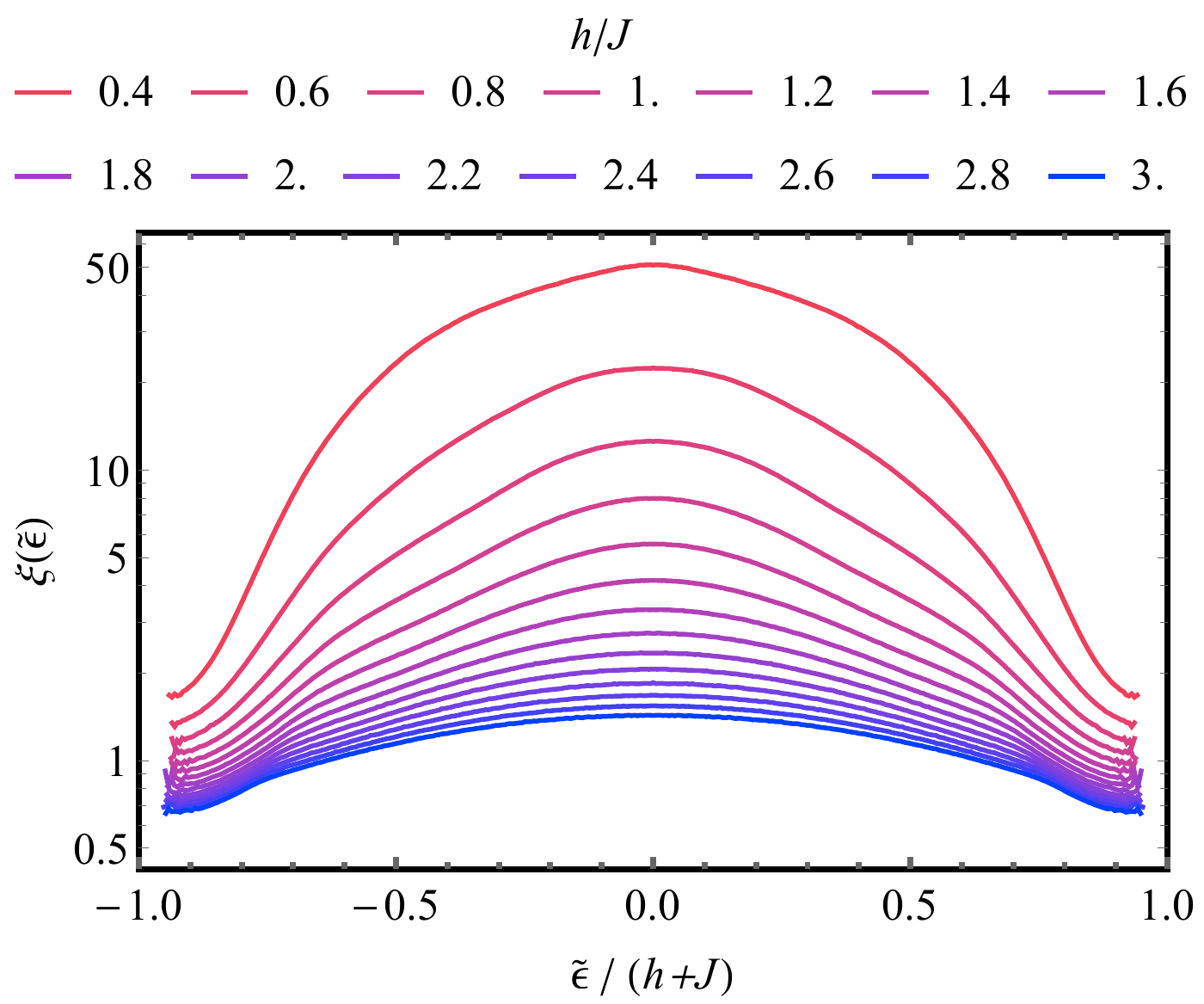}
    \caption{
    \emph{Localization length as a function of single particle energy $\tilde \epsilon$:} At any disorder strength (see legend), the localization length is maximal at $\tilde{\epsilon}=0$. We extract $\xim$ at each $h$ by averaging the localization length over a small energy window around $\tilde{\epsilon}=0$ (see Eq.~\eqref{Eq:XiMDefNum}). Data for $L = 3000$ and $4000$ samples.
    }
    \label{suppfig:xieps}

\end{figure}

\begin{figure}
    \centering
    \includegraphics[width=0.95\columnwidth]{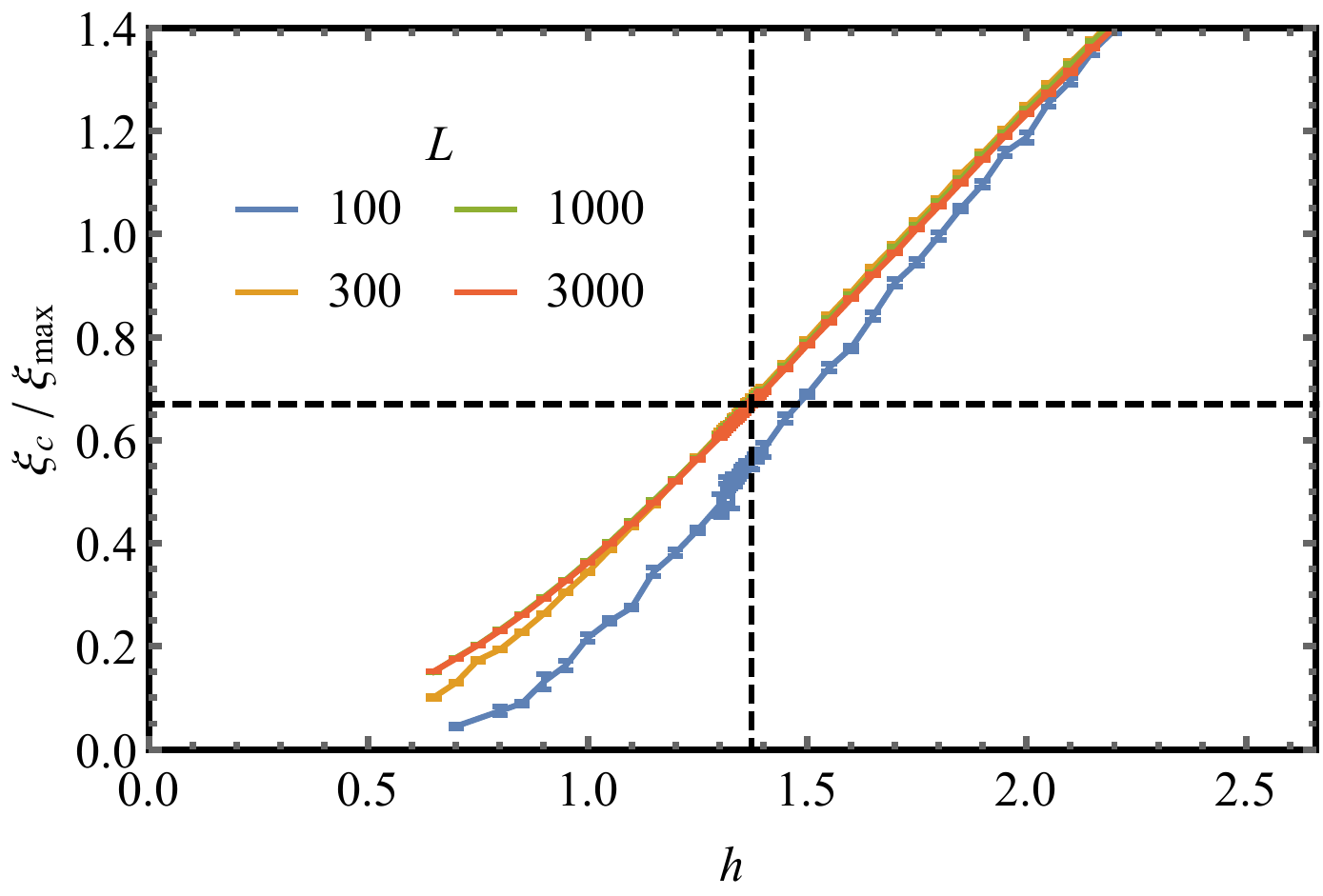}
    \caption{
    \emph{Extracting the threshold thermal fraction $\fAc$:} As the avalanche threshold is approached from the avalanching regime, $\fA \to \xic/\xim$. The plot shows the ratio $\xic/\xim$ as a function of $h$. The vertical dashed line marks $h = \hc$, at which we find $\fAc \approx 0.67 $. Data for $4000$ samples.
    }
    \label{suppfig:fc}

\end{figure}

In this section, we numerically calculate the critical disorder strength $\hc \approx 1.37$ and $\fAc \approx 0.67$ (as quoted in the main text~\eqref{eq:anderson_hc_fc}) for the Anderson model 
\begin{equation}
    J \phi^\alpha_{n+1} + J \phi^\alpha_{n-1} + 2 h_n \phi^\alpha_{n} = 2 {\tilde\epsilon}_\alpha \phi^\alpha_{n}
    \label{eq:supp_anderson}
\end{equation}
with box disorder $h_n \in [-h,h]$.

\subsection{Calculation of $\hc$}
The mean localization length $\xim$ is a monotonically decreasing function of the disorder strength. At $h = \hc $, $\xim$ equals the DRH value $\xic$. The only ingredient necessary to find $\xia$ is the distribution of localization lengths $p(\xi)$. To obtain $p(\xi)$ we numerically obtain the eigen orbitals of the Anderson model~\eqref{eq:supp_anderson} with periodic boundary conditions. The localization length of each orbital is then determined by a least squares fit to the relationship
\begin{equation}
    - \frac{|r|}{\xi_\alpha} + \mathrm{cons.}  = \log \left(\sum_{n=1}^{L} \left| \phi^\alpha_{n} \phi^\alpha_{n+r} \right| \right),
\end{equation}
where $- L/2 < r < L/2$. The distribution of values of $\xi_\alpha$ obtained from diagonalizing many such Anderson Hamiltonians provides a numerical estimate of $p(\xi)$. Example distributions of $p(\xi)$ obtained this way are shown in Fig.~\ref{suppfig:pxi}.

From these distributions, $\xia$ is easily calculated to equal $\xic$ at $\hc = 1.37\ldots$ (see Fig.~\ref{suppfig:hc}).

\subsection{Calculation of the thermal fraction of l-bits $\fA$}

To obtain $\fA$ as a function of $h$, we numerically solve for the roots of $\Delta(f)$ in Eq.~\eqref{eq:Delta}. The form of $\fA$ extracted in this way is plotted in Fig.~\ref{fig:PAval} in the main text (green solid line).

\subsection{Calculation of the critical thermal fraction $\fAc$}

As the avalanche threshold is approached from the avalanching regime, the fraction of thermal l-bits approaches the threshold value of $\fAc$:
\begin{equation}
    \lim_{h \to \hc^-} \fA = \fAc = \left. \frac{\xia}{\xim} \right|_{\xia = \xic} =  \frac{\xic}{\left. \xim \right|_{\xia = \xic}} .
\end{equation}
To obtain $\xim$ at the avalanche threshold, we use that the localization length is maximal in the centre of the single particle spectrum (see e.g. Fig~\ref{suppfig:xieps}):
\begin{equation}
    \xim = [\xi(\tilde{\epsilon} = 0)].
    \label{Eq:XiMDefNum}
\end{equation}
Numerically, the average is taken over orbitals with energies in the window $\tilde{\epsilon}_\alpha \in [-\tfrac{J+h}{100},\tfrac{J+h}{100}]$, a range significantly smaller than that over which $\xi(\epsilon)$ varies (see Fig~\ref{suppfig:xieps}).

The value of $\fAc = \left. \xia / \xim \right|_{h=\hc} = 0.67\ldots$ (dashed lines). Convergence to this value is shown for sufficiently long disordered chains ($L$ values inset) in Fig~\ref{suppfig:fc}.

\end{document}